\documentclass[preprint,showpacs,preprintnumbers,amsmath,amssymb,superscriptaddress,nofootinbib]{revtex4}
\usepackage{graphicx,color}   
\usepackage{amsmath,amssymb} 
\usepackage{url}  
\usepackage{epstopdf}
\newcommand{\hs}{\hspace*{0.5cm}} 
 
\newcommand{\be}{\begin{equation}} 
\newcommand{\ee}{\end{equation}}
\newcommand{\bea}{\begin{eqnarray}}
\newcommand{\eea}{\end{eqnarray}}
\newcommand{\nn}{\nonumber}
\newcommand{\crn}{\nonumber \\}

\newcommand{\fr}{\frac}

\newcommand{\bc}{\begin{center}}
\newcommand{\ec}{\end{center}}

\newcommand {\ba}{\begin{array}}
\newcommand {\ea}{\end{array}}
\newcommand{\ben}{\begin{enumerate}}
\newcommand{\een}{\end{enumerate}}
\newcommand{\bre}{\allowdisplaybreaks}

\usepackage{bm}
\usepackage{dcolumn}

\begin{document} 

\title{Lepton flavor violating Higgs boson decays in seesaw models: new discussions}

\author{N.H. Thao}\email{abcthao@gmail.com}
\affiliation{Department of Physics, Hanoi Pedagogical University 2, Phuc Yen, Vinh Phuc, Vietnam}
\author{L.T. Hue}\email{lthue@iop.vast.vn}
\affiliation{Institute for Research and Development, Duy Tan University, Da Nang City, Vietnam}
\affiliation{Institute of Physics,   Vietnam Academy of Science and Technology, 10 Dao Tan, Ba
 Dinh, Hanoi, Vietnam }

\author{H. T. Hung}\email{hathanhhung@hpu2.edu.vn}
\affiliation{Department of Physics, Hanoi Pedagogical University 2, Phuc Yen, Vinh Phuc, Vietnam}
\author{N.T. Xuan}\email{thixuan.ttcdd@gmail.com}
\affiliation{Department of Physics, Hanoi Pedagogical University 2, Phuc Yen, Vinh Phuc, Vietnam}

\begin{abstract}
The lepton flavor violating decay of the Standard Model-like Higgs boson (LFVHD), $h\rightarrow\mu\tau$, is discussed in  seesaw models at the one-loop level. Based on particular analytic expressions of Passarino-Veltman functions, the two  unitary and 't Hooft Feynman gauges are used to compute the branching ratio of  LFVHD and compare with  results reported recently. In the minimal seesaw (MSS) model,  the branching ratio was investigated in the whole valid range $10^{-9}-10^{15}$ GeV of new neutrino mass scale $m_{n_6}$. Using the Casas-Ibarra parameterization, this branching ratio enhances with large and increasing $m_{n_6}$. But the maximal value can reach only order of $10^{-11}$. Interesting  relations of LFVHD predicted by the MSS and inverse seesaw (ISS) model are discussed. The ratio between two LFVHD branching ratios predicted by the ISS and MSS is simply $m^2_{n_6}\mu^{-2}_X$, where $\mu_X$ is the small neutrino mass scale in the ISS. The consistence between different calculations is shown precisely from  analytical approach. 
\end{abstract}
\pacs{
12.15.Lk, 12.60.-i, 13.15.+g,  14.60.St 
 }
\maketitle
\section{\label{intro} Introduction}  
After the Higgs boson was observed by ATLAS and CMS \cite{higgsdicovery}, the LFVHD has been searched experimentally \cite{exLFVh}, where upper bounds for branching ratios (\textbf{Brs}) of the decays $h\rightarrow\mu\tau, e\tau$ are order of $\mathcal{O}(10^{-2})$. Signals of LFVHD at future colliders have been discussed, where sensitivities for detecting these channel decays are shown to be $10^{-5}$ in the near future  \cite{LFVcol}. Up to now, the lepton flavor violating (LFV) decays  of the standard-model-like and new Higgs bosons have been investigated in many models beyond the standard model (SM) \cite{Apo,Apo1,EArganda,iseesaw,e1612.09290,moress,LFVgeneral,SUSY,THDL2,LFVHDUgauge,KHHung}.  Among them, the MSS \cite{MSS} is the simplest that can explain successfully the recent neutrino data. Naturally, the mixing between different flavor neutrinos leads to many LFV processes from loop corrections. But it predicts very suppressed branching ratios (Br) of LFV decays of charged leptons. Recent studies on the Br of LFVHD were also shown to be very small \cite{EArganda}. In contrast, the ISS \cite{ISSmodel}, another simple extension of the SM, predicts much larger values of LFV branching ratios, including those of LFVHD \cite{iseesaw,e1612.09290}. In fact, the Br of LFVHD in the ISS  were calculated in many different ways in order to guarantee the consistence of the LFVHD amplitudes. 

We stress that understanding the  mechanism for generating loop corrections to \textbf{Brs}  of LFVHD in simple models like the MSS and ISS is very important for studying LFVHD processes in other complicated models. That is why LFVHD predicted by these two models were discussed in many works, for example  \cite{Apo,Apo1,EArganda,iseesaw,e1612.09290,moress}. In the ISS, recent results in \cite{iseesaw} showed that branching ratios of LFVHD increase with increasing values of very heavy neutrino masses when  the Casas-Ibarra method \cite{ibarra} was applied to formulating the Yukawa couplings of heavy neutrinos \footnote{We thank Dr. E. Arganda for this comment}.  But the Brs are always constrained by upper bounds  because of the perturbative limit of the Yukawa couplings. Using the mass insertion approximation, a recent study \cite{e1612.09290} also calculated the Br of LFVHD in the ISS model in both unitary and 't Hooft Feynman, where  previous results in \cite{iseesaw} were confirmed to be well consistent in the region of parameters containing large new neutrino mass scale $m_{n_6}$.  The above discussions indicate that although  one-loop contributions   in both MSS and ISS arise from the same set of Feynman diagrams, the two models predict very different Br values. The reason is the appearance of a small mass scale $\mu_X$  in the ISS, which gives tiny contributions to  the heavy neutrino masses, but affects strongly on the neutrino mixing matrix. Hence there should exist  simple relations between two expressions of Brs predicted by the two models. These interesting relations were not discussed previously, therefore will be focused in this work. We will show that if  $m_{n_6}$ is large enough,  the ratio between Brs of LFVHD  of the ISS and MSS is order of $ m^2_{n_6}\mu^{-2}_{X}$, enough to explain clearly the LFVHD difference between two models.

Regarding the MSS,  LFVHD was  discussed mainly in  ranges of $10^2-10^7$  GeV  \cite{Apo,EArganda}, while the valid range  of the new neutrino mass scale is from $\mathcal{O}(10^{-9})$ GeV to $\mathcal{O}(10^{15})$ GeV. In addition, a good estimation made in Ref. \cite{Apo} suggested that the Br may enhance with increasing masses of heavy neutrinos, even when  the Casas-Ibarra parameterization is used. We note that this parameterization are now still widely used to investigate the signal of seesaw models  at recent colliders \cite{SSbound}. As a result,  possibilities that large Brs  of LFVHD may exist in ranges of new neutrino mass scales that were not mentioned previously. Therefore,  studies the LFVHD  in the whole valid range as well  as new approaches  to compare well-known results and confirm  consistent  analytic formulas for calculating Br of LFVHD in seesaw models are still interesting and necessary. These are main scopes  of this work.  In particular, in order to  guarantee the stability  of numerical results at very large values of $m_{n_6}$,   LFVHD  processes will be computed using  analytic expressions of  Passarino-Veltman functions (PV functions) given in ref. \cite{LFVHDUgauge}.  Using a mathematica code based on these functions, we found that it is much easier and more  convenient to increase the precision  than using available  numerical packages such as  Looptools \cite{looptool}.
 This makes our calculation different from all previous works.  In addition, the one-loop contributions to LFVHD in both unitary and 't Hooft Feynman gauges will be constructed using notations in \cite{LFVHDUgauge}. Then we  cross-check the consistence between total amplitudes calculated in two gauges, and  the ones established in previous works  \cite{Apo,EArganda,iseesaw}. A detailed checking divergence cancellation will be presented analytically. For the MSS, after showing that Br of LFVHD is suppressed with small $m_{n_6}$, we will pay attention mainly to the region with large $m_{n_6}$. To guarantee the consistence of our investigation on LFVHD in the MSS, the connection between analytic formulas of  LFVHD amplitudes in the two models MSS and ISS will be discussed deeply. In this work,  Yukawa couplings of new neutrinos are only investigated following the Casas-Ibarra parameterization  \cite{ibarra}. This parameterization was used to investigate independently LFVHD processes predicted by the MSS and ISS in Refs. \cite{EArganda,iseesaw}, where other important properties of LFVHD were presented in details. 

Our work is arranged as follows. Sec. \ref{general}  establishes notations and couplings of a general seesaw model needed for studying LFVHD. In Sec. \ref{Amp}, we  construct LFVHD amplitudes in two unitary and 't Hooft Feynman gauges using notations of PV functions given in \cite{LFVHDUgauge}. Then we prove the divergent cancellation and the consistence between two expressions of the LFVHD amplitudes. In  Sec. \ref{LFVHD}, we show the choice of parameterizing the  neutrino mixing matrices. After that, the Brs of LFVHD  are numerically investigated. We will focus  on new results of LFVHD in the MSS, and interesting relations between the Brs predicted by two models MSS and ISS. Sec. \ref{Con} summarizes new results of  this work. 
\section{\label{general}General formalism and couplings for LFVHD}
 The general seesaw model  is different from the Standard Model (SM) by $K$ additional right-handed neutrinos, $N_{R,I}\sim (1,1,0)$ with $I=1,2,...,K$  \cite{numixing}.  The new  Lagrangian part is 
\be -\Delta \mathcal{L}=Y_{\nu,aI}\overline{\psi_{L,a}}\widetilde{\phi} N_{R,I} + \frac{1}{2}\overline{(N_{R,I})^{c}} m_{M,IJ}N_{R,J} + \mathrm{h.c.},\label{SSterm}\ee
where $a=1,2,3$; I,J=1,2,...,K;   $\psi_{L,a}=(\nu_{L,a},e_{L,a})^T$ are $SU(2)_L$ lepton doublets    and $(N_{R,I})^c=C\overline{N_{R,I}}^T$. The Higgs bosons are also doublets  $\phi=(G_W^+, (h+iG_Z+v)/\sqrt{2})^T$ and  $\widetilde{\phi}=i\sigma_2\phi^*$.  Each of them consists of  three   Goldstone bosons of $W^\pm$ and $Z$ bosons; a neutral CP-even Higgs boson $h$ and  the vacuum expectation value (VEV), $\langle\phi\rangle=\frac{v}{\sqrt{2}}=174$ GeV ($v=246 $ GeV).  Notations for flavor states of active neutrinos are $ \nu_L=(\nu_{L,1},\;\nu_{L,2},\;\nu_{L,3})^T$ and $ (\nu_L)^c\equiv ((\nu_{L,1})^c,\;(\nu_{L,2})^c,\;(\nu_{L,3})^c)^T$.  Notations for new neutrinos are  $ N_R=(N_{R,1},\; N_{R,2},...,\;N_{R,K})^T$, and $ (N_R)^c=((N_{R,1})^c,\; (N_{R,2})^c,...,\;(N_{R,K})^c)^T$. In the bases of the original neutrinos,  $\nu'_{L}\equiv(\nu_L,\;(N_R)^c )^T$ and $(\nu'_L)^c=((\nu_L)^c,\;N_R)^T$, the Lagrangian part (\ref{SSterm}) generates  the  following mass term for neutrinos,
\bea - \mathcal{L}^{\nu}_{\mathrm{mass}}
    \equiv\frac{1}{2}\overline{\nu'_L}
    M^{\nu}(\nu'_L)^c+ \mathrm{h.c.} 
    =\frac{1}{2}\overline{\nu'_L}
    \left(
\begin{array}{cc}
0 & M_D \\
 M_D^T & M_{N} \\
  \end{array}
  \right) (\nu'_L)^c+ \mathrm{h.c.},
\label{L0numass} \eea 
where $M_N$ is a  symmetric and non-singular $K\times K$ matrix, and $M_D$ is a $3\times K$ matrix, $(M_D)_{aI}=Y_{\nu,aI}\langle\phi\rangle$. The matrix $M^{\nu}$ is symmetric, therefore it can be diagonalized via $(K+3)\times(K+3)$  matrix, $U^\nu$, satisfying the unitary condition,  $U^{\nu\dagger}U^{\nu}=I$.  We define
\bea U^{\nu T}M^{\nu}U^{\nu}=\hat{M}^{\nu}=\mathrm{diagonal}(m_{n_1},m_{n_2},m_{n_3}, m_{n_4},..., m_{n_{(K+3)}}), \label{diaMnu} \eea
where   $m_{n_i}$ ($i=1,2,...,K+3$) are  mass eigenvalues of the $(K+3)$ mass eigenstates $n_{L,i}$, i.e. physical states of  neutrinos. Three  light active neutrinos are $n_{L,a}$  with $a=1,2,3$. 
The relation between the flavor  and mass eigenstates are
\bea\nu'_L=U^{\nu*} n_L, \hs \mathrm{and} \; (\nu'_L)^c=U^{\nu}  (n_L)^c, \label{Nutrans}
 \eea
 where $n_L\equiv(n_{L,1},n_{L,2},...,n_{L,K+3})^T$.

 In  calculation, we will use a general notation of four-component (Dirac) spinor, $n_i$ ($i=1,2,..,K+3$), for all active and exotic neutrinos. Specifically, a  Majorana fermion $n_i$ is  defined as   $n_i\equiv(n_{L,i},\; (n_{L,i})^c)^T=n^c_i=(n_i)^c$. The chiral components are $n_{L,i}\equiv P_{L}n_i$ and $n_{R,i}\equiv P_{R}n_i= (n_{L,i})^c$, where $P_{L,R}=\frac{1\pm\gamma_5}{2}$ are chiral operators.  The similar definitions  for the original neutrino states  are  $\nu_a \equiv (\nu_{L,a},\; (\nu_{L,a})^c)^T$, $N_I\equiv((N_{R,I})^c,\; N_{R,I})^T$, and $\nu'=(\nu,\,N)^T$.  The relations in (\ref{Nutrans}) are rewritten as follows, 
 \be P_L\nu'_i=\nu'_{L,i} =U^{\nu*}_{ij}n_{L,j},\; \mathrm{and}\; P_R\nu'_i=\nu'_{R,i} =U^{\nu}_{ij}n_{R,j}, \hs i,j=1,2,...,K+3, \label{Nutrans2}\ee
 where more precise expressions are  $\nu_{L,a}= P_L\nu'_a = U^{\nu*}_{ai}n_{L,i}$, $(N_{R,I})^c= P_L\nu'_{I+3} = U^{\nu*}_{(I+3)j}n_{L,j}$, $(\nu_{L,a})^c= P_R\nu'_a = U^{\nu}_{ai}n_{R,i}$, and $N_{R,I}= P_R\nu'_{I+3} = U^{\nu}_{(I+3)j}n_{R,j}$  ($I=1,2,3,..,K$).

As usual, the covariant derivative is  $D_{\mu}=\partial_{\mu}-ig T^aW^a-ig'YB_{\mu}$. We emphasize that the signs in $D_{\mu}$ will result in signs of couplings $hG^{\pm}_WW^{\pm}$ and $\overline{e}_a\nu_a W^-$. Correspondingly, the lepton flavor violating (LFV) couplings of $W^\pm$ boson to leptons are,
\bea \mathcal{L}^{\mathrm{lep}}_{\mathrm{kin}}= i\overline{\psi_{L,a}}\gamma^{\mu}D_{\mu}\psi_{L,a}
&\supset&  \frac{g}{\sqrt{2}}\left( \overline{\nu_{L,a}}\gamma^{\mu} e_{L,a} W^+_{\mu}+ \overline{ e_{L,a}}\gamma^{\mu}\nu_{L,a} W^-_{\mu}\right) \crn
&=&\frac{g}{\sqrt{2}}\left(U^{\nu}_{aj} \overline{n_{j}}\gamma^{\mu}P_L e_{a} W^+_{\mu}+ U^{\nu*}_{aj}  \overline{ e_{a}}\gamma^{\mu}P_Ln_{j} W^-_{\mu}\right), \label{wnue}  \eea
where $a=1,2,3$; and $j=1,2,...,K+3$.  

The Yukawa couplings  that contribute to LFVHD are
\bre
\bea - \mathcal{L}^{\mathrm{lep}}_{\mathrm{Y}} &=& y_{e_a} \overline{\psi_{L,a}}\phi e_{R,a}+ Y_{\nu,aI}\overline{\psi_{L,a}}\widetilde{\phi} N_{R,I}+ \mathrm{h.c.}\crn
&\supset&  \frac{m_{e_a}}{v} h \overline{e_a}e_a  + \frac{\sqrt{2}m_{e_a}}{v}\left(U^{\nu}_{aj} G^+_{W} \overline{n_{L,j}}e_{R,a} + U^{\nu*}_{aj}G^-_{W}\overline{e_{R,a}}n_{L,j} \right)\crn
&+&Y_{\nu,aI}\left[-G^-_{W}\overline{e_{L,a}}N_{R,I}- G^+_{W} \overline{N_{R,I}}e_{L,a}\right]\crn
&+& \frac{1}{v\sqrt{2}}h \left[(M_D)_{aI}  \overline{\nu_{L,a}} N_{R,I}  +(M_D)^*_{aI}\overline{ N_{R,I}}\nu_{L,a} \right].
 \label{lfvYukwas}\eea
Using $(M_D)_{aI}=M^{\nu}_{a(I+3)}$, and $N_{R,I}=\nu'_{R,(I+3)}$, the last line in (\ref{lfvYukwas}) changes in to the new form,
   $\frac{1}{v} h \overline{n_{i}}\left[M^{\nu}_{a(I+3)}U^{\nu}_{ai}U^{\nu}_{(I+3)j}P_R+M^{\nu*}_{a(I+3)} U^{\nu*}_{(I+3)i}U^{\nu*}_{aj} P_L \right] n_j$.  
  It can be proved that  
 \be
M^{\nu}_{a(I+3)}U^{\nu}_{ai}U^{\nu}_{(I+3)j}P_R+M^{\nu*}_{a(I+3)} U^{\nu*}_{(I+3)i}U^{\nu*}_{aj} P_L= \left(\sum_{a=1}^3U^{\nu}_{ai}U^{\nu*}_{aj}\right)\left(m_{n_i}P_L+m_{n_j}P_R\right),\label{nutr2}
\ee
which was given in \cite{EArganda,iseesaw}. A proof is as follows, based on the following properties of  $M^{\nu}$ and $U^{\nu}$ defined in Eqs.  (\ref{L0numass}) and (\ref{diaMnu}), 
\bea
M^{\nu}_{ab}&=&0,\hs M^{\nu}_{(I+3)(J+3)}=(m_N)_{IJ},\hs M^{\nu}_{a(I+3)}=(M_D)_{aI}, \hs M^{\nu}_{(I+3)a}=(M_D^T)_{Ia},\crn
U^{\nu \dagger}U^{\nu}&=&I,\hs M^{\nu}= U^{\nu*}\hat{M}^{\nu}U^{\nu\dagger}, \hs \mathrm{and} \; M^{\nu*}= U^{\nu}\hat{M}^{\nu}U^{\nu T}.  \label{property1} \eea
 The first term in the left hand side of  Eq. (\ref{nutr2}) will change  exactly into the second term in the right hand side of Eq. (\ref{nutr2}),  after  mediate steps of transformation, namely  
  \bea
M^{\nu}_{a(I+3)}U^{\nu}_{ai}U^{\nu}_{(I+3)j}&=&  \left(U^{\nu*}\hat{M}^{\nu} U^{\nu\dagger}\right)_{a(I+3)}U^{\nu}_{ai}U^{\nu}_{(I+3)j}= U^{\nu*}_{ak}m_{n_k} U^{\nu\dagger}_{k(I+3)}U^{\nu}_{ai}U^{\nu}_{(I+3)j} \crn
&=&  U^{\nu*}_{ak}U^{\nu}_{ai}m_{n_k}\left(\sum_{l=1}^{K+3} U^{\nu\dagger}_{kl}U^{\nu}_{lj}- \sum_{b=1}^{3}U^{\nu\dagger}_{kb}U^{\nu}_{bj} \right) =
 U^{\nu*}_{ak}U^{\nu}_{ai}m_{\nu_k}\left( \delta_{kj} -U^{\nu\dagger}_{kb}U^{\nu}_{bj} \right) \crn
 &=&U^{\nu*}_{aj}U^{\nu}_{ai}m_{n_j}- U^{\nu}_{ai}U^{\nu}_{bj} \left(U^{\nu*}_{ak}m_{n_k}U^{\nu\dagger}_{kb}\right) =U^{\nu*}_{aj}U^{\nu}_{ai}m_{n_j}- U^{\nu}_{ai}U^{\nu}_{bj}M^{\nu*}_{ab} \crn
 &=&U^{\nu}_{ai} U^{\nu*}_{aj}m_{n_j}.  \label{t1nutr2}
\eea
   From (\ref{t1nutr2}), the second term in the left hand side of (\ref{nutr2}) can be derived easily,  $ M^{\nu*}_{a(I+3)} U^{\nu*}_{(I+3)i}U^{\nu*}_{aj}= \left[M^{\nu}_{a(I+3)}U^{\nu}_{aj} U^{\nu}_{(I+3)i}\right]^*= \left[U^{\nu}_{aj} U^{\nu*}_{ai}m_{n_i}\right]^*
= U^{\nu}_{ai} U^{\nu*}_{aj} m_{n_i}$. Finally, the Feynman rule for the vertex (\ref{nutr2}) with two Majorana leptons $h\overline{n_i}n_j$ must be expressed in a symmetric form {\footnote{ We thank Dr. E. Arganda for showing us this point}}, namely 
$  -\frac{g}{4m_W}\sum_{i,j}\overline{n_i} \left[\left(m_{n_i} C_{ij} +m_{n_j} C^*_{ij}\right)P_L + \left(m_{n_j} C_{ij} +m_{n_i} C^*_{ij}\right)P_R \right]n_j,$
where $C_{ij}=\sum_{c=1}^3U^{\nu}_{ci} U^{\nu*}_{cj}$  \cite{Apo,spinor} .

The couplings relating with $G^\pm_{W}$ are proved the same way, namely
\bea  Y_{\nu,aI}\overline{e_{L,a}}N_{R,I} G^-_{W}&=& \frac{\sqrt{2}}{v} (M_D)_{aI}\overline{e_{L,a}}N_{R,I} G^-_{W}
=\frac{g}{\sqrt{2}m_W} U^{\nu*}_{ai}\overline{e_{a}}P_Rn_{i} G^-_{W}.\nn\eea 
The vertices relating to LFVHD are collected in Table \ref{lfvcoupling}.
\begin{table}[h]
  \centering
  \begin{tabular}{|c|c|c|c|}
  \hline
  Vertex & coupling & Vertex & coupling \\
   \hline
 $h W^{+\mu} W^{-\nu}$  & $igm_Wg_{\mu\nu}$  & $hG^{+}_WG^{-}_W$  &  $-\frac{ig m^2_h}{2 m_W}$\\
 \hline
 $h G^{+}_W W^{-\mu}$  &$\frac{ig}{2}\left( p_{+}-p_0\right)_{\mu}$   & $hG^{-}_WW^{+\mu}$  &   $\frac{ig}{2}\left( p_{0}-p_-\right)_{\mu}$\\
  \hline
  $\overline{n_{i}}e_{a}W^+_{\mu}$  & $ \frac{ig}{\sqrt{2}}U^{\nu}_{ai} \gamma^{\mu}P_L$  & $\overline{ e_{a}}n_{i} W^-_{\mu}$ &$\frac{ig}{\sqrt{2}} U^{\nu*}_{ai}\gamma^{\mu}P_L$  \\
  \hline
  $\overline{n_{i}}e_{a} G^+_{W}$  & $ -\frac{ig}{\sqrt{2} m_W}U^{\nu}_{ai}\left(m_{e_a}P_R- m_{n_i}P_L\right)$  & $\overline{ e_{a}}n_{i} G^-_{W}$ &$ -\frac{ig}{\sqrt{2} m_W}U^{\nu*}_{ai}\left( m_{e_a}P_L- m_{n_i}P_R\right)$  \\
  \hline
  $ h  \overline{n_{i}} n_j $  & $ \frac{-ig}{2m_W}\left[ C_{ij}\left(P_Lm_{n_i}+ P_R m_{n_j}\right)\right.$ &$h\overline{e_a}e_a$  &  $- \frac{igm_{e_a}}{2 m_W}$\\
    & $\quad\quad\left. +C^*_{ij}\left(P_Lm_{n_j}+ P_R m_{n_i}\right)\right]$ & &  \\
  \hline
\end{tabular}
  \caption{Couplings relating with LFVHD  in seesaw models. Here, $C_{ij}=\sum_{c=1}^3U^{\nu}_{ci} U^{\nu*}_{cj}$. The $p_0,\, p_+$ and $p_-$ are  incoming momenta of  $h$, $G^+_{W}$ and $G^{-}_W$, respectively.}\label{lfvcoupling}
\end{table}
We note that the coupling $hG^+_{W}G^-_{W}$ in Table \ref{lfvcoupling} is consistent with that given in \cite{hgwgw,e1612.09290}.

The effective Lagrangian of  the LFVHD  is  written as
$\mathcal{L}^{LFV}= h \left(\Delta_L \overline{\mu}P_L \tau +\Delta_R \overline{\mu}P_R \tau\right) + \mathrm{h.c.}$,
where   $\Delta_{L,R}$ are scalar factors arising from loop contributions.
The partial decay width  is
\be
\Gamma (h\rightarrow \mu\tau)\equiv\Gamma (h\rightarrow \mu^{-} \tau^{+})+\Gamma (h\rightarrow \mu^{+} \tau^{-})
\simeq  \fr{ m_{h} }{8\pi }\left(\vert \Delta_L\vert^2+\vert \Delta_R\vert^2\right), \label{LFVwidth}
\ee
where  $m_{h}\gg m_2,m_3$ and $m_2,m_3$ being  masses of muon and tau, respectively. The on-shell conditions for external momenta are $p^2_{a}=m_a^2$ ($a=2,3$) and $ p_h^2 \equiv( p_2+p_3)^2=m^2_{h}$, $m_h=125$ GeV. Next,  $\Delta_{L,R}$ with be calculated at one-loop level, in two gauges of unitary and 't Hooft Feynman.
\section{\label{Amp}Analytic amplitudes and divergence cancellation}
\subsection{Amplitude in the unitary gauge and divergence cancellation}
\begin{figure}[h]
	\centering
	\includegraphics[width=15cm]{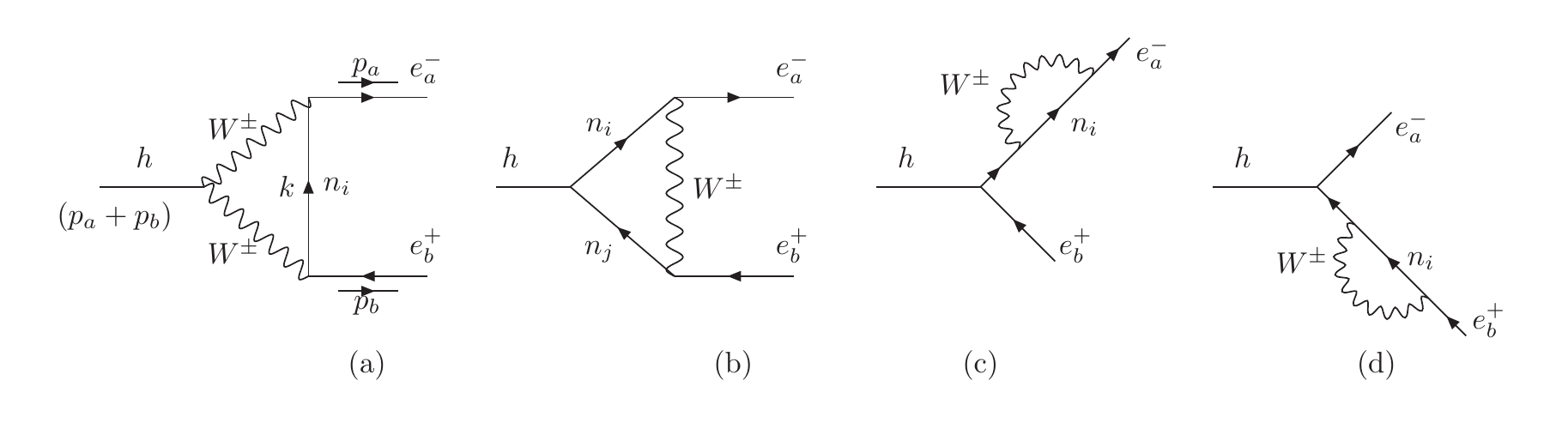}\\
	\caption{Feynman diagrams contributing to LFVHD in the unitary gauge. }\label{Feydia1}
\end{figure}

In the unitary gauge, the Feynman diagrams for a decay $h\rightarrow e^-_ae^+_b$ $(a<b)$ are presented in Fig. \ref{Feydia1}.
 The  loop contributions are written as $\Delta_{L,R}= \Delta^{(a)}_{L,R}+ \Delta^{(b)}_{L,R}+ \Delta^{(c+d)}_{L,R}$, where  the three terms come from private contributions of diagrams \ref{Feydia1}a), \ref{Feydia1}b), and  sum of contributions from two diagrams c) and d),  respectively.  The analytic expressions of contributions from the three  diagrams \ref{Feydia1}a), c),  and d) can be derived directly from  \cite{LFVHDUgauge}, except the  diagram \ref{Feydia1}b) containing the coupling $h\overline{n_i}n_j$. An  analytic expression of  $ \Delta^{(b)}_{L,R}$ is derived in  appendix \ref{Diagramb}. We have used Form \cite{form} to cross-check our results. In addition, the total $\Delta_{L,R}$ is consistent with the result calculated in the 't Hooft Feynman gauge, as we will show later.  Expressions of LFVHD contributions in the unitary gauge  are 
{\small 
	\bea \Delta^{(a)}_L&=& -\frac{g^3 m_a}{64\pi^2 m_W^3}\sum_{i=1}^{K+3}U^{\nu*}_{ai}U^{\nu}_{bi}
	\left\{ m_{n_i}^2\left(B^{(1)}_1- B^{(1)}_0- B^{(2)}_0\right) -m_b^2 B^{(2)}_1  +\left(2m_W^2+m^2_{h}\right)m_{n_i}^2 C_0 \right.\crn &-&\left. \left[2m_W^2\left(2m_W^2+m_{n_i}^2+m_a^2-m_b^2\right) + m_{n_i}^2m_h^2\right] C_1 +
	\left[2m_W^2\left(m_a^2-m^2_{h}\right)+ m_b^2 m^2_{h}\right]C_2\frac{}{}\right\},\crn
	\Delta^{(a)}_R&=& -\frac{g^3 m_b}{64\pi^2 m_W^3}\sum_{i=1}^{K+3}U^{\nu*}_{ai}U^{\nu}_{bi}
	\left\{ -m_{n_i}^2\left(B^{(2)}_1+B^{(1)}_0+ B^{(2)}_0\right) +m_a^2 B^{(1)}_1  +\left(2m_W^2+m^2_{h}\right)m_{n_i}^2 C_0 \right.\crn &-&\left. 
	\left[2m_W^2\left(m_b^2-m^2_{h}\right)+ m_a^2 m^2_{h}\right]C_1 + \left[2m_W^2\left(2m_W^2+m_{n_i}^2-m_a^2+m_b^2\right) + m_{n_i}^2m_h^2\right] C_2 \frac{}{}\right\},	
	\label{dela}\crn
	\Delta^{(b)}_L  &=& - \frac{g^3  m_a}{64\pi^2m^3_W}\sum_{i,j=1}^{K+3}U^{\nu*}_{ai}U^{\nu}_{bj} \left\{ C_{ij}
	\left[  m^2_{n_i}B^{(1)}_1 +m^2_{n_j}B^{(12)}_0-m^2_{n_j}m^2_W C_0
	\right.\right.\crn
	&+&\left.\left.   \left[ 2m^2_{n_i} m^2_{n_j} +2m_W^2\left(m^2_{n_i}+m^2_{n_j}\right) -(m^2_{n_i}m^2_b+m^2_{n_j}m^2_a)\right]C_1 \right]\right.\crn
	&+& \left. C^*_{ij} m_{n_i}m_{n_j}\left[ B^{(12)}_0+ B^{(1)}_1 -m_W^2C_0 +\left(4 m_W^2+m^2_{n_i} +m^2_{n_j}-m_a^2-m_b^2 \right)C_1\right]
	 \right\}, \crn
	\Delta^{(b)}_R  &=& - \frac{g^3  m_b}{64\pi^2m^3_W}\sum_{i,j=1}^{K+3}U^{\nu*}_{ai}U^{\nu}_{bj} \left\{C_{ij}
	\left[ - m^2_{n_j}B^{(2)}_1 +m^2_{n_i}B^{(12)}_0-m^2_{n_i}m^2_W C_0
	\right.\right.\crn
	&-&\left.\left.  \left[2m^2_{n_i} m^2_{n_j} +2m_W^2(m^2_{n_i}+m^2_{n_j})- (m^2_{n_i}m^2_b+m^2_{n_j}m^2_a) \right]C_2 \right]\right.\crn
	&+&\left. C^*_{ij} m_{n_i}m_{n_j}\left[ B^{(12)}_0-B^{(2)}_1 -m_W^2C_0 -\left(4 m_W^2+m^2_{n_i} +m^2_{n_j}-m_a^2-m_b^2 \right)C_2\right] \right\},
	\label{delb}\eea
}
and
\bea \Delta^{(c+d)}_{L} &=& \frac{g^3m_a}{64\pi^2m^3_W} \sum_{i=1}^{K+3}U^{\nu*}_{ai}U^{\nu}_{bi}\fr{m_b^2}{(m_a^2-m_b^2)}\left[\left(2 m_W^2 +m_{n_i}^2\right) \left(B^{(1)}_1 +B^{(2)}_1 \right) \right. \crn&+&\left. m_a^2 B^{(1)}_1 +m_b^2 B^{(2)}_1 -  2m_{n_i}^2\left(B^{(1)}_0-B^{(2)}_0\right)\right],  \label{DfvL} \\
\Delta^{(c+d)}_{R}&=& \frac{m_a}{m_b}\Delta^{(c+d)}_{L}.   \label{delcd}\eea
Regarding $\Delta^{(b)}_{L,R}$,  the contributions from $ B^{(1)}_1=B^{(1)}_1(m_W^2,m^2_{n_i})$ and $B^{(2)}_1$ are zeros because, for example,  $B^{(1)}_1$ contains a factor $\sum_{j} U^{\nu}_{bj}m_{n_j}U^{\nu}_{cj}=\left(U^{\nu*}\hat{M}^{\nu}U^{\nu\dagger}\right)^*_{bc}= M^{\nu*}_{bc}=0$.

Divergence cancellation in the total amplitude is explained as follows. From divergent parts of the PV functions in Appendix \ref{PVfunction}, the divergent parts of $\Delta^{(a)}_{L}$ and $\Delta^{(b)}_{L}$ are 
\bea  \mathrm{Div}[\Delta^{(a)}_L] &=&-  \frac{g^3m_a}{64\pi^2m^3_W}\sum_{i=1}^{K+3}U^{\nu*}_{ai}U^{\nu}_{bi}\left[m^2_{n_i}\left( \frac{-3}{2}\Delta_\epsilon\right)+ m^2_b\frac{1}{2}\Delta_\epsilon\right]  \crn
&=& \frac{3g^3m_a}{128\pi^2 m_W^3}\Delta_\epsilon\sum_{i=1}^{K+3}U^{\nu*}_{ai}U^{\nu}_{bi}m^2_{n_i} ,\crn
\mathrm{Div}[\Delta^{(b)}_L] &=& -  \frac{g^3m_a}{64\pi^2m^3_W}\left[ \sum_{i,j=1}^{K+3}\sum_{c=1}^{3}U^{\nu*}_{ai}U^{\nu}_{ci}U^{\nu*}_{cj}U^{\nu}_{bj} \left(m^2_{n_i}\frac{1} {2}\Delta_\epsilon+m^2_{n_j}\Delta_\epsilon\right)\right.  \crn 
&+&\left. \sum_{i,j=1}^{K+3}\sum_{c=1}^{3}U^{\nu*}_{ai}U^{\nu*}_{ci}U^{\nu}_{cj}U^{\nu}_{bj} m_{n_i}m_{n_j}\Delta_\epsilon \right]
\crn
&=&  \frac{g^3 m_a}{128\pi^2m^3_W}\Delta_\epsilon\left[\sum_{i,j=1}^{K+3} \sum_{c=1}^{3}U^{\nu*}_{ai}U^{\nu}_{ci}U^{\nu*}_{cj}U^{\nu}_{bj}\left(m^2_{n_i}+2m^2_{n_j} \right)\right.\crn
&+&\left. \sum_{i,j=1}^{K+3}\sum_{c=1}^{3}U^{\nu*}_{ai}U^{\nu*}_{ci}U^{\nu}_{cj}U^{\nu}_{bj} 2m_{n_i}m_{n_j} \right], \label{divdelab} \eea
where the unitary property of $U^{\nu}$ is used to cancel the second term of $\mathrm{Div}[\Delta^{(a)}_L]$, namely $ \sum_{i=1}^{K+3}U^{\nu*}_{ai}U^{\nu}_{bi}=\left(U^{\nu}U^{\nu\dagger}\right)_{ab}=0$.  The second term of  $\mathrm{Div}[\Delta^{(b)}_L]$ vanishes because $ \sum_{i}U^{\nu*}_{ai}U^{\nu*}_{ci}  m_{n_i}=\left(U^{\nu*}\hat{M}_{\nu}U^{\nu\dagger}\right)_{ac}=M^{\nu}_{ac}=0$ with all $a,c=1,2,3$. We simplify the first term of  $\mathrm{Div}[\Delta^{(b)}_L]$ based on the following equalities
\bea
\sum_{i,j=1}^{K+3}\sum_{c=1}^{3}m^2_{n_i}U^{\nu*}_{ai}U^{\nu}_{ci}U^{\nu*}_{cj}U^{\nu}_{bj} &=&\sum_{i=1}^{K+3}\sum_{c=1}^{3}m^2_{n_i}U^{\nu*}_{ai}U^{\nu}_{ci}\sum_{j=1}^{K+3}U^{\nu*}_{cj}U^{\nu}_{bj}\crn
&=&\sum_{i=1}^{K+3}\sum_{c=1}^{3}m^2_{n_i}U^{\nu*}_{ai}U^{\nu}_{ci}(U^{\nu}U^{\nu\dagger})_{bc} =\sum_{i=1}^{K+3}m^2_{n_i}U^{\nu*}_{ai}U^{\nu}_{bi}.\label{CancelCij}
\eea
Similarly, we have $\sum_{i,j=1}^{K+3}\sum_{c=1}^{3}2m^2_{n_j}U^{\nu*}_{aj}U^{\nu}_{ci}U^{\nu*}_{cj}U^{\nu}_{bj}= \sum_{i=1}^{K+3}2m^2_{n_i}U^{\nu*}_{ai}U^{\nu}_{bi}$.  Inserting these two results into  $\mathrm{Div}[\Delta^{(b)}_L] $  will give $\mathrm{Div}[\Delta^{(b)}_L]+\mathrm{Div}[\Delta^{(a)}_L]=0$. With  $\Delta^{(c+d)}_L$, the  divergent parts of the two terms $m_a^2 B^{(1)}_1$ and  $m_b^2 B^{(2)}_1$ vanish because of the GIM mechanism, while two  sums $ [B^{(1)}_1 +B^{(2)}_1]$ and $ [B^{(1)}_0 -B^{(2)}_0]$ are finite. Hence,  $\Delta_L$ is finite.   $\Delta_R$ has the same conclusion.  

\subsection{Amplitude in the 't Hooft Feynman gauge.}
In the 't Hooft Feynman gauge, there are ten form factors $F^{(i)}_{L,R}$, ($i=1,2,..,10$) corresponding to ten diagrams shown in Fig. 1 of Refs. \cite{EArganda,iseesaw}. The total contribution is $\Delta_{L,R}=\sum_{i=1}^{10} F^i_{L,R}$. Formulas of $F^{(i)}_{L,R}$ in terms of PV functions defined in \cite{LFVHDUgauge} are as follows, 
{\small
	\bea F^{(1)}_L &=&-\frac{g^3 m_a}{64\pi^2m_W^3} \sum_{i,j=1}^{K+3} B_{ai}B^*_{bj}\left\{C_{ij}\left[ m^2_{n_j}\left(B^{(12)}_0 +m^2_WC_0\right) -\left(m_a^2m^2_{n_j}+ m_b^2m^2_{n_i}-2m_{n_i}^2m^2_{n_j}  \right)C_1 \right]\right. \crn
	&+&\left. m_{n_i}m_{n_j} C^*_{ij}\left[ B^{(12)}_0 +m^2_WC_0 +\left(m^2_{n_i}+ m^2_{n_j}-m^2_a-m^2_b\right)C_1 \right] \right\},\crn
	F^{(1)}_R &=& -\frac{g^3m_b}{64\pi^2m_W^3} \sum_{i,j=1}^{K+3} B_{ai}B^*_{bj}\left\{C_{ij}\left[ m^2_{n_i}\left(B^{(12)}_0 +m^2_WC_0\right) +\left(m_a^2m^2_{n_j}+ m_b^2m^2_{n_i}-2m_{n_i}^2m^2_{n_j}  \right)C_2 \right]\right. \crn
	&+&\left. m_{n_i}m_{n_j} C^*_{ij}\left[B^{(12)}_0 +m^2_WC_0 -\left(m^2_{n_i}+ m^2_{n_j}-m^2_a-m^2_b\right)C_2 \right] \right\},\crn  
	F^{(2)}_L &=& \frac{g^3 m_a}{64\pi^2m_W^3} \sum_{i,j=1}^{K+3} B_{ai}B^*_{bj}\times 2 m_W^2 \crn
	&\times&\left\{C_{ij}\left[m^2_{n_j}C_0- \left( m^2_{n_i}+m^2_{n_j} \right)C_1 \right] + m_{n_i}m_{n_j}C^*_{ij}\left(C_0-2C_1\right)\right\},\crn
	F^{(2)}_R &=&  \frac{g^3m_b}{64 \pi^2m_W^3}  \sum_{i,j=1}^{K+3} B_{ai}B^*_{bj}\times 2 m_W^2 \crn
	&\times&\left\{C_{ij}\left[m^2_{n_i}C_0+ \left( m^2_{n_i}+m^2_{n_j} \right)C_2 \right]+ m_{n_i}m_{n_j}C^*_{ij}\left(C_0+2C_2\right)\right\},  \label{f12}\\
	F^{(3)}_L &=& \frac{g^3m_a}{64\pi^2 m_W^3} \sum_{i=1}^{K+3} B_{ai}B^*_{bi}\left[4 m_W^4C_1\right], \hs
	F^{(3)}_R =   \frac{g^3m_b}{64\pi^2m_W^3} \sum_{i=1}^{K+3} B_{ai}B^*_{bi}\left[-4 m_W^4C_2\right],\crn
	F^{(4)}_L &=& -\frac{g^3m_a}{64\pi^2m_W^3} \sum_{i=1}^{K+3} B_{ai}B^*_{bi}\times m_W^2\left[- m_{n_i}^2 C_0 +\left(2 m_b^2-m^2_{n_i}\right) C_1 -m^2_b C_2\right],\crn
	F^{(4)}_R &=& - \frac{g^3 m_b}{64\pi^2 m_W^3}
	\sum_{i=1}^{K+3} B_{ai}B^*_{bi}m_W^2 \left[B^{(12)}_0 +3m^2_{n_i} C_0 +\left(2m^2_{h} -2m^2_b -m^2_a \right)C_1 +\left(m^2_{n_i} +2m^2_b\right)C_2 \right], \crn 
	F^{(5)}_L &=& -\frac{g^3m_a}{64\pi^2 m_W^3} \sum_{i=1}^{K+3} B_{ai}B^*_{bi}m_W^2 \left[B^{(12)}_0 +3m^2_{n_i} C_0 -\left(m^2_{n_i} +2m^2_a\right)C_1  -\left(2m^2_{h} -m^2_b -2m^2_a \right)C_2\right], \crn
	F^{(5)}_R &=& -\frac{g^3m_b}{64\pi^2m_W^3} \sum_{i=1}^{K+3} B_{ai}B^*_{bi} m_W^2\left[ -m_{n_i}^2 C_0 +m^2_a C_1 -\left(2 m_a^2 -m^2_{n_i}\right) C_2 \right], \crn
	F^{(6)}_L &=&- \frac{g^3m_a}{64\pi^2 m^3_W} \sum_{i=1}^{K+3} B_{ai}B^*_{bi}\times m_h^2\left[ m^2_{n_i}(C_0-C_1) +m^2_bC_2\right],\crn
	F^{(6)}_R &=& -\frac{g^3m_b}{64\pi^2 m^3_W} \sum_{i=1}^{K+3} B_{ai}B^*_{bi}\times m_h^2\left[ m^2_{n_i}(C_0+C_2) -m^2_aC_1\right], \label{f3456}\\
	F^{(7)}_L &=&\frac{g^3m_a}{64\pi^2m_W^3} \sum_{i=1}^{K+3} B_{ai}B^*_{bi} \frac{(D-2)m_W^2m_b^2}{(m_a^2-m_b^2)}B^{(1)}_1, \hs F^{(7)}_R = \frac{m_a}{m_b}F^{(7)}_L,\crn
	F^{(9)}_L &=&\frac{g^3m_a}{64\pi^2m_W^2} \sum_{i=1}^{K+3} B_{ai}B^*_{bi} \frac{(D-2)m_W^2m^2_b}{(m_a^2-m_b^2)}B^{(2)}_1, \hs F^{(9)}_R =\frac{m_a}{m_b}F^{(9)}_L,
	\label{f79}\\
	F^{(8)}_L &=& -\frac{g^3m_a}{64\pi^2m^3_W} \sum_{i=1}^{K+3} B_{ai}B^*_{bi} \frac{m_b^2}{(m_a^2-m_b^2)}\left[2 m^2_{n_i}B^{(1)}_0  -\left( m^2_{n_i}+m_a^2\right)B^{(1)}_1\right],\crn
	F^{(8)}_R &=& -\frac{g^3m_b}{64\pi^2m^3_W}  \sum_{i=1}^{K+3} B_{ai}B^*_{bi}\frac{1}{(m_a^2-m_b^2)} \left[m^2_{n_i}\left(m_a^2+m_b^2\right) B^{(1)}_0 -m_a^2\left( m^2_{n_i}+m_b^2\right)B^{(1)}_1\right],\crn
	F^{(10)}_L &=& \frac{g^3m_a}{64\pi^2m^3_W}  \sum_{i=1}^{K+3} B_{ai}B^*_{bi} \frac{1}{(m_a^2-m_b^2)} \left[m^2_{n_i}\left(m_a^2+m_b^2\right) B^{(2)}_0  +m_b^2\left( m^2_{n_i}+m_a^2\right)B^{(2)}_1\right],\crn
	F^{(10)}_R &=& \frac{g^3m_b}{64\pi^2m^3_W} \sum_{i=1}^{K+3} B_{ai}B^*_{bi} \frac{m^2_a}{(m_a^2-m_b^2)}\left[ 2 m^2_{n_i}B^{(2)}_0 +\left(m^2_{n_i}+m_b^2\right)B^{(2)}_1 \right], \label{f810}\eea
}
 where $B_{ai}=U^{\nu*}_{ai}, B^*_{bj}=U^{\nu}_{bj}$,  $C_{ij}=\sum_{c=1}^3U^{\nu}_{ci}U^{\nu*}_{cj}$, and $D=4-2\epsilon$ is the integral dimension defined in Appendix \ref{PVfunction}.  Although $F^{(7)}_{L,R}$ and $F^{(9)}_{L,R}$ contain $B$-functions, they are finite because of the GIM mechanism. Hence it can be replaced with $D=4$.  
 Because  $B^{(12)}_0=B^{(12)}_0(m_W^2,m_W^2)$ in $F^{(4)}_R$ and $F^{(5)}_L$  do not depend on $m_{n_i}$, therefore  vanish because of the GIM mechanism. They will be ignored from now on.   

Although our notations of PV functions are different from those  in \cite{EArganda,iseesaw},  transformations  between two sets of notations are, (see a detailed proving in  Appendix \ref{match}) 
\bea C_{0} & \leftrightarrow& C_0, \hs C_{1} \leftrightarrow C_{12}-C_{11},  \hs C_2  \leftrightarrow C_{12}, \crn
B^{(12)}_{0} &\leftrightarrow&   B_{0}(m^2_W,m^2_W),\; B_{0}(m^2_{n_i},m^2_{n_j}), \quad  B^{(1,2)}_{0}(M^2_0, M^2) \leftrightarrow B_{0}(m^2_{l_{k,m}},M^2_0, M^2), \crn
B^{(1)}_{1}(M^2_0, M^2)&\leftrightarrow& -   B_{1}(m^2_{l_k}, M^2_0, M^2), \hs B^{(2)}_{1}(M^2_0, M^2)\leftrightarrow  B_{1}(m^2_{l_m}, M^2_0, M^2). \label{twonotation}\eea
The PV functions used in our work were checked to be consistent with Looptools \cite{looptool}, see details in \cite{KHHung}. The differences between our results and those shown in \cite{iseesaw} are minus signs in  $F^{(4)}_{L,R}$ and $F^{(5)}_{L,R}$. Our formulas are consistent with the results presented in Ref.  \cite{e1612.09290}{\footnote{The correct Feynman rule for the coupling $h\overline{n_i} n_j$ gives  consistent $F^{(1,2)}_{L,R}$  with those in Ref. \cite{iseesaw}.}}, where the authors  confirmed that these  signs do not affect the results given in Ref. \cite{iseesaw}.

 Now we will  check the consistence between total amplitudes calculated in two gauges. Regarding to triangle diagrams with two internal neutrino lines,  the deviation of contributions in two gauge are determined as follows,  
\bea \delta_1&=& \Delta^{(b)}_L-\left(F^{(1)}_L+F^{(2)}_L \right)= -\frac{g^3}{4m_W^3}\frac{m_a}{16\pi^2} \sum_{i,j=1}^{K+3} B_{ai}B^*_{bj}C_{ij}m^2_{n_i}B^{(1)}_1 (m^2_{W},m^2_{n_i})
 \crn &=&-  \frac{g^3}{4m_W^3}\frac{m_a}{16\pi^2} \sum_{i=1}^{K+3} B_{ai}B^*_{bi}m^2_{n_i}\left( B^{(1)}_0 (m^2_{n_i},m^2_{W})- B^{(1)}_1 (m^2_{n_i},m^2_{W})\right), \label{del1}\eea
where  useful equalities of B-functions are used  \cite{bardin}. In  addition, $C_{ij}$ in the first line of (\ref{del1}) is simplified using the same trick given in (\ref{CancelCij}). 
Similarly, other deviations are
\bea \delta_2&=& \Delta^{(a)}_L- \sum^{6}_{k=3}F^{(k)}_L=  -\frac{g^3}{4m_W^3}\frac{m_a}{16\pi^2}\sum_{i=1}^{K+3} B_{ai}B^*_{bi} \left[-m_b^2 B^{(2)}_1 -m^2_{n_i}\left(B^{(1)}_0 -B^{(1)}_1 +B^{(2)}_0\right)\right],\crn
\delta_3&=& \Delta^{(c+d)}_L-\sum^{10}_{k=7}F^{(k)}_L=  -\frac{g^3}{4m_W^3}\frac{m_a}{16\pi^2} \sum_{i=1}^{K+3} B_{ai}B^*_{bi} \left[ m_b^2 B^{(2)}_1 + m^2_{n_i}B^{(2)}_0\right], \label{dev23}\eea
where $B_{0,1,2}\equiv B_{0,1,2} (m^2_{n_i},m^2_{W})$. 
Then, it can be seen easily that $\delta_1 +\delta_2 +\delta_3=0$. Hence,  the  total amplitudes calculated in two gauges are the same. 

\section{\label{LFVHD} LFVHD in the minimal and inverse seesaw models} 
\subsection{Parameterization the neutrino mixing matrix}
To start, we  consider a general expression of the neutrino mixing matrix $U^{\nu}$  \cite{numixing}, 
 \be U^{\nu}= \Omega \left(
                      \begin{array}{cc}
                        U & \mathbf{O} \\
                        \mathbf{O} & V \\
                      \end{array}
                    \right), \hs
  \label{Unuform}\ee
 where  $\mathbf{O}$ is a  $3\times K$ null matrix,  $U$ and  $V$ are   $3\times3$ and $K\times K$ unitary matrices, respectively. The $\Omega$ is a $(K+3)\times (K+3)$ unitary matrix that can be formally written as
 \be \Omega=\exp\left(
                  \begin{array}{cc}
                    \mathbf{O} & R \\
                    -R^\dagger & \mathbf{O} \\
                  \end{array}
                \right)=
 \left(
  \begin{array}{cc}
                    1-\frac{1}{2}RR^{\dagger} & R \\
                    -R^\dagger &  1-\frac{1}{2}R^{\dagger} R\\
                  \end{array}
                \right)+ \mathcal{O}(R^3),
    \label{Ommatrix}\ee
where $R$ is a $3\times K$ matrix where  absolute values of al elements are smaller than unity. The unitary matrix  $ U=U_{\mathrm{PMNS}}$ is the Pontecorvo-Maki-Nakagawa-Sakata (PMNS) matrix \cite{upmns}.

The mass matrices of  neutrinos are written as follows, 
\bea  \hat{M}_N&=&\mathrm{diag}(m_{n_4},\;m_{n_5},...,\;m_{n_{K+3}}),\crn
m_\nu&=&U^*_{\mathrm{PMNS}} \mathrm{diag}(m_{n_1},\;m_{n_2},\;m_{n_3}) U^{\dagger}_{\mathrm{PMNS}}=U^*_{\mathrm{PMNS}} \hat{m}_{\nu}U^{\dagger}_{\mathrm{PMNS}}, \label{newnote}\eea
where $m_{n_i}$ is the physical masses of all neutrinos, 
\bea U_{\mathrm{PMNS}}=\left(
\begin{array}{ccc}
	c_{12}c_{13} & s_{12}c_{13} & s_{13} e^{-i\delta} \\
	-s_{12}c_{23}-c_{12}s_{23}s_{13}e^{i\delta} & c_{12}c_{23}-s_{12}s_{23}s_{13}e^{i\delta} & s_{23}c_{13} \\
	s_{12}s_{23}-c_{12}c_{23}s_{13}e^{i\delta} & -c_{12}s_{23}-s_{12}c_{23}s_{13}e^{i\delta} & c_{23}c_{13} \\
\end{array}
\right) \mathrm{diag}(1,\; e^{i\alpha},\;e^{i\beta}),
\label{umns}\eea
and $c_{ab}\equiv\cos\theta_{ab}$, $s_{ab}\equiv\sin\theta_{ab}$. In the normal hierarchy scheme, the best-fit  values of  neutrino oscillation parameters are given as  \cite{actnuUpdate}{\footnote{ Updated neutrino data can be found in \cite{pdg2016}.  But our main results are unchanged}}
\bea \Delta m^2_{21}&=& 7.50\times 10^{-5}\;\mathrm{ eV^2},\hs  \Delta m^2_{31}= 2.457\times 10^{-3}\; \mathrm{eV^2},\crn
s^2_{12}&=&0.304,\; s^2_{23}=0.452,\; s^2_{13}=0.0218, \label{nuosc}\eea
where $ \Delta m^2_{a1}=m^2_{n_a}-m^2_{n_1}$ ($a=2,3$). In this work, other parameters will be fixed as $\delta=\alpha=\beta=0$. 

The condition of seesaw mechanism for neutrino mass generation is  $|M_D|\ll |M_N|$, where $|M_D|$ and  $|M_N|$ denote characteristic scales of $M_D$ and $M_N$,  resulting in useful relations {\footnote{We thank LE Duc Ninh for pointing out factors $1/2$ in the last relation in (\ref{masafla}).}} \cite{numixing},
 \bea   R^* &\simeq& M_{D}M_N^{-1}, \hs m_{\nu}\simeq-M_DM_N^{-1} M^T_D, \crn
  V^* \hat{M}_N V^{\dagger}&\simeq& M_N+ \frac{1}{2}R^TR^* M_N+ \frac{1}{2} M_NR^{\dagger} R. \label{masafla}\eea 
Based on the second relation in (\ref{masafla}), the  matrix $M_D$ can be parameterized via a general $K\times3$ matrix $\xi$, which satisfies the only condition  $ \xi^T\xi=I_3$ \cite{ibarra,numixing,EArganda}, namely  
\be M_D^T=iU_N^{*}\left(M^d_N\right)^{1/2}\xi \left(\hat{m}_{\nu}\right)^{1/2}U^{\dagger}_{\mathrm{PMNS}},\label{fmD}\ee
where  $U_N$ is an  unitary matrix diagonalizing $M_N$, $U_N^TM_NU_N=M^d_N=\mathrm{diag}(M_1,M_2,...,M_K)$.  

In the MSS mentioned in \cite{Apo,EArganda}, the particle content is different from the Standard Model (SM) by three additional right-handed neutrinos ($K=3$), $N_{R,I}\sim (1,1,0)$ with $I=1,2,3$.  New notations of neutrino mass matrices are $ m_D\equiv M_D$, and $m_M\equiv M_N$. They are the respective $3\times3$ Dirac and Majorana mass matrices  corresponding to  the first and second term of (\ref{SSterm}), $(m_D)_{iJ}=Y_{\nu,iJ}\langle\phi\rangle$, and $(m_M)_{iJ}=m_{M,iJ}$. The matrix $m_{M}$ is real, symmetric and non-singular. 

The mixing matrix in the ISS model considered in ref. \cite{iseesaw} can be found approximately using the above general discussion with $K=6$.  Relations of  notations between two parameterizations in  \cite{iseesaw} and \cite{numixing} are  
\bea
 M_D= (m_D,\hs \mathcal{O}), \hs M_N=\left(
                                      \begin{array}{cc}
                                        \mathcal{O}& M_R \\
                                        M^T_R & \mu_X \\
                                      \end{array}
                                    \right), \hs m_{\nu}=M_{\mathrm{light}},
  \label{repara}
\eea
where $\mathcal{O}$ is  the $3\times3$ matrix with all elements being zeros. From the definition of the inverse matrix,  $M^{-1}_NM_N=M_NM^{-1}_N=I_6$, we derive that
\bea  M^{-1}_N= \left(
                  \begin{array}{cc}
                    -M^{-1} & \left(M^T_R\right)^{-1} \\
                    M_R^{-1} & 0 \\
                  \end{array}
                \right),
\label{invMN}\eea
where $M$ is defined as 
$M=M_R\mu_X^{-1}M_R^T$ \cite{iseesaw}.
 From (\ref{masafla}), we then find that \cite{numixing}
\bea R^*= M_D M_N^{-1} &=& \left(-m_D M^{-1},\hs m_D\left(M^T_R\right)^{-1} \right), \crn
 m_{\nu}=-M_DM_N^{-1}M_D^T &=& m_D\left( M_R^T\right)^{-1}\mu_XM_R^{-1}m_D^T=m_D M^{-1}m_D^T.
 \label{R1}\eea 
 These two expressions  are consistent with those given in \cite{iseesaw,numixing}, giving a parameterization of  $m_D$  as follows, 
 \be m^T_D=U_M^{*}\mathrm{diag}(\sqrt{M_1},\; \sqrt{M_2},\; \sqrt{M_3}) \xi'\sqrt{\hat{m}_\nu}  U^{\dagger}_{\mathrm{PMNS}},\label{mDiss} \ee
 where $U_M$ satisfies $M=U_M^{*}\mathrm{diag}(M_1,\; M_2,\; M_3)U^{\dagger}_M$ and $\xi'$ is a complex orthogonal matrix satisfying $\xi'\xi'^T=I_3$. The mixing matrix $U^\nu$ now is a $9\times9$ matrix.

In order to compare and mark relations between LFVHD in two MSS and ISS models, we will pay attention to only simply cases of  choosing parameters.  In the  MSS model, the choice is $\xi=U_N=I_3$, leading to  following  simple expressions of Eqs.  in  (\ref{masafla}), namely 
\be M^d_N=M_N,\quad  R = -iU_{\mathrm{PMNS}}\,\hat{m}_{\nu}^{1/2} \left(M^d_N\right)^{-1/2}, \hs 
V=I_3, \quad \hat{M}_N= M^d_N+ \hat{m}_{\nu}.\label{masafla1}\ee

In the ISS model, from (\ref{mDiss}) we see that $m_D$ is parameterized in terms of many free parameters, hence it is enough to choose that $\mu_X=\mu_X I_3$. This parameter is a new scale making the most important difference  between  the neutrino mixing  matrices in the ISS and MSS. We also assume that   $M_R=\hat{M}_R=\mathrm{diag}(M_{R_1},\,M_{R_2},\,M_{R_3})$ and $\xi'=I_3$. With $|\mu_X|\ll |M_R|$ we have 
\bea U_M=I_3,\quad M^d_N= \left(\begin{matrix}
	\hat{M}_R	& 0 \\ 
	0	& \hat{M}_R
\end{matrix} \right), \qquad V\simeq \dfrac{1}{\sqrt{2}}
\left(\begin{matrix}
	-iI_3	& I_3 \\ 
	iI_3	& I_3
\end{matrix} \right). \label{UNiss}\eea
We can see that both $\hat{M}_R$ (ISS) and $M_N$ (MSS) play roles as  exotic neutrino mass scales.  Therefore, they are identified as neutrino masses in both models, $\hat{M}_R=M_N=\mathrm{diag}(m_{n_4},\, m_{n_5},\,m_{n_6})$. The differences between two models now are two mixing matrix $V$ in (\ref{UNiss}) and $R$,  and the $\mu_X$ scale, which does not appear in the MSS model. The $\mu_X$ plays special roles in the ISS model via its appearance in the second sub-matrix of the mixing matrix $R$ given in (\ref{R1}).  A simple relation  between largest elements of  $R$ matrices in two models is
\be R^{\mathrm{ISS}} \sim \sqrt{\frac{m_{n_6}}{\mu_X}} R^{\mathrm{MSS}},\label{Rss}\ee
where $m_{n_6}$ now is considered as exotic neutrino mass scale, $m_{n_4}\le m_{n_5}\le m_{n_6}$. The relation (\ref{Rss}) is the main reason that explains why the Br of LFVHD predicted from the ISS is much larger than that from the MSS.   

In the following, we will discuss on LFVHD in  the MSS  model. The results of LFVHD in the ISS model can be derived from discussion  in the MSS model based on  (\ref{Rss}). 
\subsection{\label{Dis}Discussion on LFVHD}
\begin{figure}[ht] 
	\centering 
	\begin{tabular}{cc}
		\includegraphics[width=7.5cm]{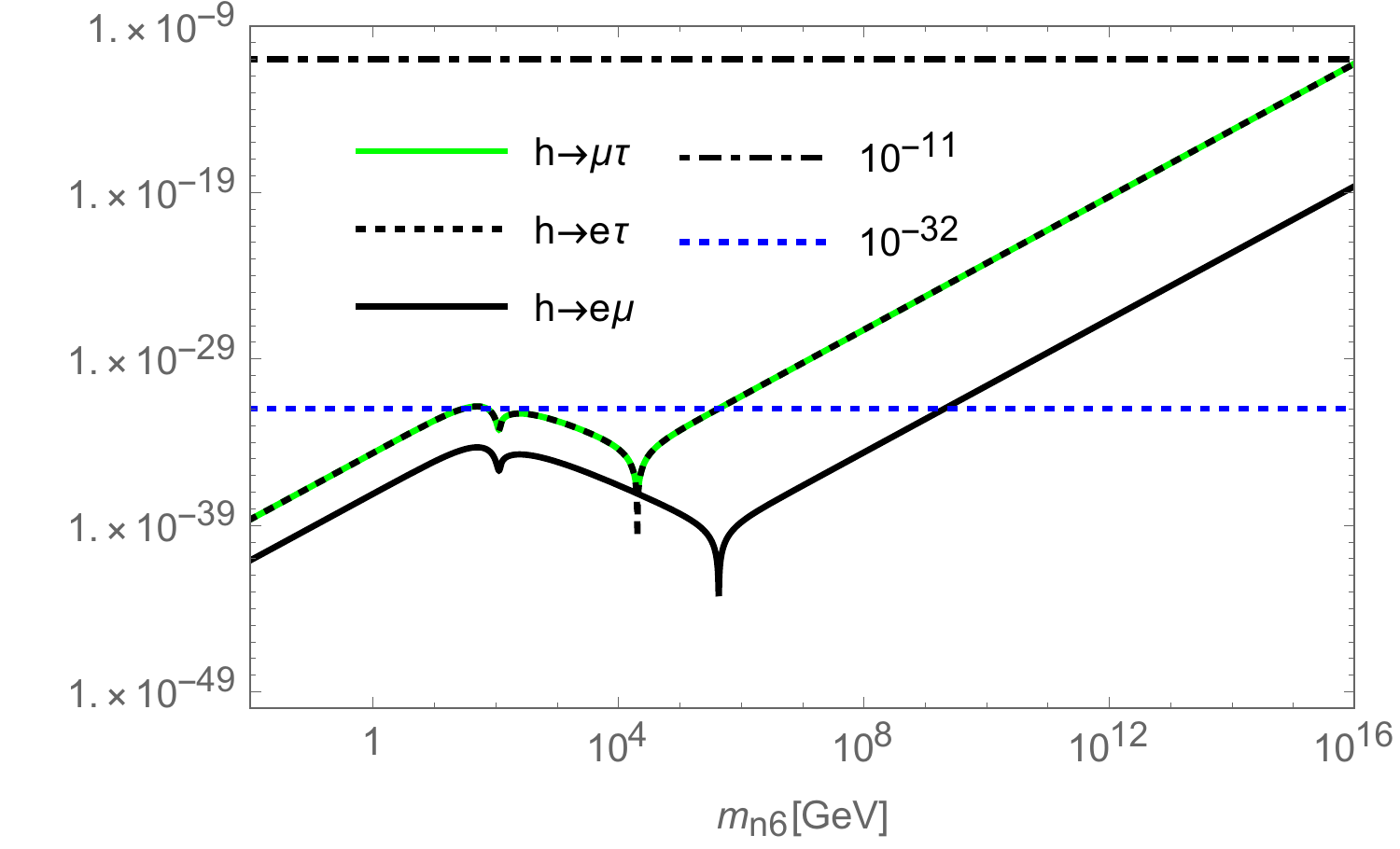}& \includegraphics[width=7.5cm]{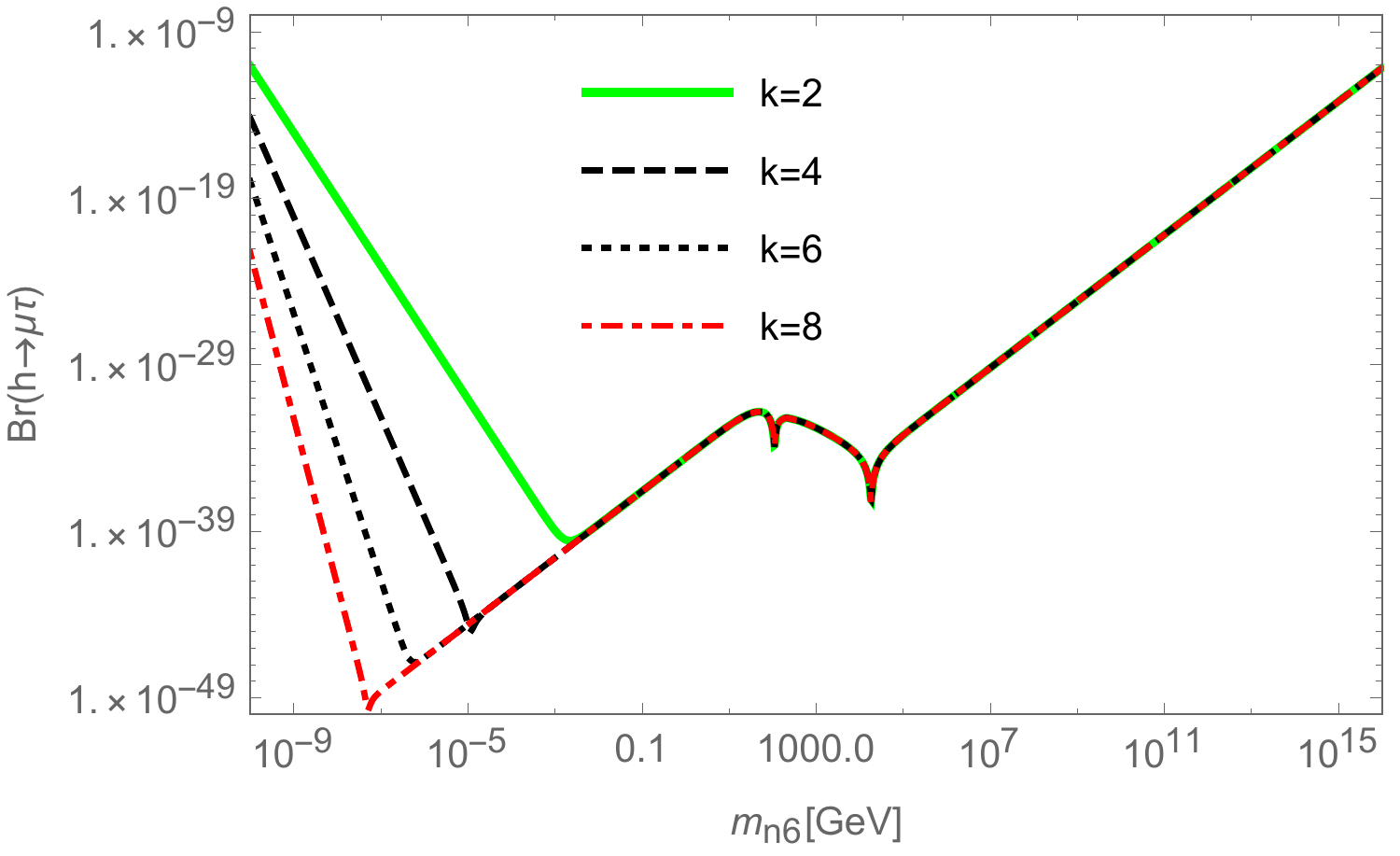}\\
	\end{tabular}
	\caption{Left panel: Br$(h\rightarrow e_ae_b)$ as functions of  $m_{n_6}$ with  non-degenerate heavy neutrino masses. Right panel:  The dependence of Br$(h\rightarrow e_ae_b)$ on the mixing matrix $U^{\nu}$  up to an order $\mathcal{O}(R^k)$ with $k=2,4,6,8$.}\label{NBrhmutau1}
\end{figure}

In the MSS model, our investigation will use three physical masses of exotic neutrinos, $m_{n_{4,5,6}}$,  as free parameters. The matrix $M_D$ can be derived from relations (\ref{fmD}),  i.e $M_D=iU^*_{\mathrm{PMNS}}\left(M^d_N\,\hat{m}_{\nu}\right)^{1/2}$.  As a result, the mixing matrix $U^{\nu}$ is written as a function of physical neutrino masses and $U_{\mathrm{PMNS}}$. 
To determine constrains  of heavy neutrino masses $m_{n_6}$, we base on relations in (\ref{masafla}), which suggest that $m_{n_6}\times m_{n_3}\simeq |M_D|^2< 6\pi\times 174^2$, because of the perturbative limit of the Yukawa couplings $Y_{\nu, ij}$ \cite{iseesaw}. Combing with the  active neutrino data given in (\ref{nuosc}), where at least one active neutrino mass is not smaller than $\sqrt{\Delta m_{31}^2}=5\times 10^{-11}$ GeV, we get an upper constrain, $m_{n_6}< 8\times 10^{15}$ GeV, when $m_{n_1}\ll\sqrt{\Delta m_{31}^2}$.  The lower constrain is $m_{n_6}> |M_{D}|> m_{n_3}>5\times 10^{-11}$ GeV.  Numerical illustrations are shown in Fig. \ref{NBrhmutau1}, where  three  heavy neutrino masses are non-degenerate, $3m_{n_4}=2m_{n_5}=m_{n_6}$,  and  $m_{n_1}=10^{-12}\,\mathrm{GeV}\ll\sqrt{ \Delta m_{31}^2}$.  

The left panel  of Fig. \ref{NBrhmutau1}  presents  Br$(h\rightarrow e_{a}e_{b})$ as functions of $m_{n_6}$. Unlike previous works such as \cite{Apo,EArganda}, heavy neutrinos masses were not considered at the interesting scale above  $10^{10}$ GeV, where leptogenesis can be successful explained in the MSS frame work \cite{leptoMSS}. More important,  large values of heavy neutrinos may give large Br of LFVHD, as we have seen numerically.   Unfortunately, values of  $m_{n_6}\le 8\times 10^{15}$ GeV gives an upper bound Br$(h\rightarrow \mu\tau)\le \mathcal{O}(10^{-11})$.  For other two decays, we get the relations Br$(h\rightarrow e\tau)\simeq$Br$(h\rightarrow \mu\tau)=(m^2_{\tau}/m^2_{\mu})$Br$(h\rightarrow e\mu)\simeq 287\times\mathrm{Br}(h\rightarrow e\mu)$.  Hence, we just focus on the Br$(h\rightarrow\mu\tau)$. 

The right panel of Fig. \ref{NBrhmutau1} shows values of Br$(h\rightarrow\mu\tau)$ in the whole valid range of $m_{n_6}$, namely $10^{-10}<m_{n_6}<8\times 10^{15}$ [GeV], where $U^{\nu}$ is considered up to $\mathcal{O}(R^k)$. Each curve  separates into three different parts. In the part with very heavy exotic neutrino masses,  $m^2_{n_6}\gg m^2_h,m^2_{W}$, i.e. $m_{n_6}>\mathcal{O}\left(10^{4}\right)$, we found a simple relation: Br$(h\rightarrow\mu\tau)= 6.3\times 10^{-44}m_{n_6}^2\,[\mathrm{GeV}^2]$. On the other hand, for the part with very small  exotic neutrino masses,  $m^2_{n_6}\ll m^2_\mu,m^2_{\tau}$, i.e.  $m_{n_6}<\mathcal{O}\left(10^{-3}\right)$, there appears a new relation: Br$(h\rightarrow\mu\tau)=\dfrac{8.7\times 10^{-52}}{(m_{n_6}\mathrm{GeV})^4}$, when the matrix $\Omega$ is calculated up to $\mathcal{O}(R^2)$. This will lead to the maximal values of Br$(h\rightarrow\mu\tau)\le 10^{-11}$, the same order with large $m_{n_6}\sim \mathcal{O}(10^{15})$ GeV. If the matrix $\Omega$ is calculated more exactly, the Br$(h\rightarrow\mu\tau)$ will decrease significantly with small $m_{n_6}$, but will not change with large $m_{n_6}$. This can be explained from the conditions of the matrix $\Omega$, which is written in terms of the power series in $R$. If $m_{n_6}$ is small, $R\sim \sqrt{|m_{\nu}|/m_{n_6}}$ will be large as $m_{n_6}\rightarrow |M_D|\rightarrow |m_{\nu}|$. The calculation will be less accurate with smaller power $k$ included in $\Omega$. We consider  more cases of $U^{\nu}$ where the matrix $\Omega$ in (\ref{Ommatrix}) is considered up to order $\mathcal{O}(R^8)$.  We conclude that the Br$(h\rightarrow \mu\tau)$ is very suppressed with small masses of exotic neutrinos. In contrast, large $m_{n_6}$ results in $|R|\ll 1$. Therefore, it is enough to consider the mixing matrix $U^{\nu}$ with order of $\mathcal{O}(R^2)$  in the region where $m_{n_6}\ge 0.1$ GeV. In conclusion, to find large Br$(h\rightarrow\mu\tau)$, we just consider the region with  large $m_{n_6}$.

To explain why large Br$(h\rightarrow\mu\tau)$ corresponds to large $m_{n_6}$, we pay attention to the properties of the mixing matrix $U^{\nu}$, the PV-functions and factors relating with them in the  expressions of $\Delta^{(a)}_{L,R}$, $\Delta^{(b)}_{L,R}$, and $\Delta^{(c+d)}_{L,R}$. When $m^2_{n_I}\gg m^2_h,m^2_W$, the terms with factors $m^2_{n_{I}}$ will give dominant contributions. The PV functions containing  $m^2_{n_{I}}$ will have the following properties:  $B_{0,1,2}(m^2_{n_i})=\mathcal{O}(10)$, $C_{0,1,2}(m^2_{n_{i}})\sim \ln(m^2_{n_6})/m^2_{n_6}$. Hence the largest contributions will come from $m^2_{n_6}B_{0,1,2}\sim m^2_{n_6} $ in $\Delta^{(a+c+d,b)}_{L,R}$ and $m^4_{n_6} C_{0,1,2}\sim [\ln\,m^2_{n_6}]m^2_{n_6}$ in $\Delta^{(b)}_{L,R}$.
\begin{figure}[ht]
	\centering 
	\begin{tabular}{cc}
		\includegraphics[width=7.5cm]{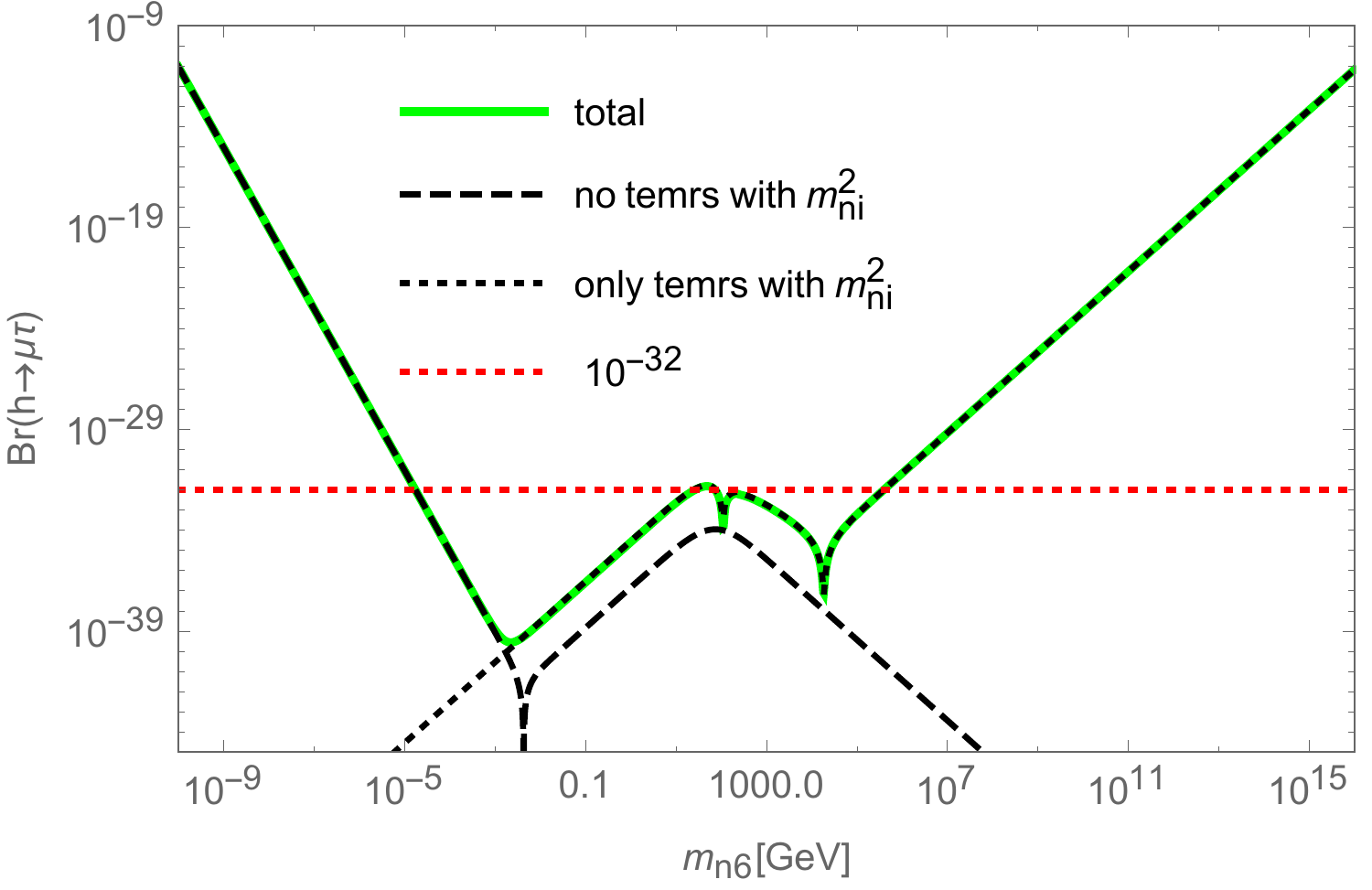}& \includegraphics[width=7.5cm]{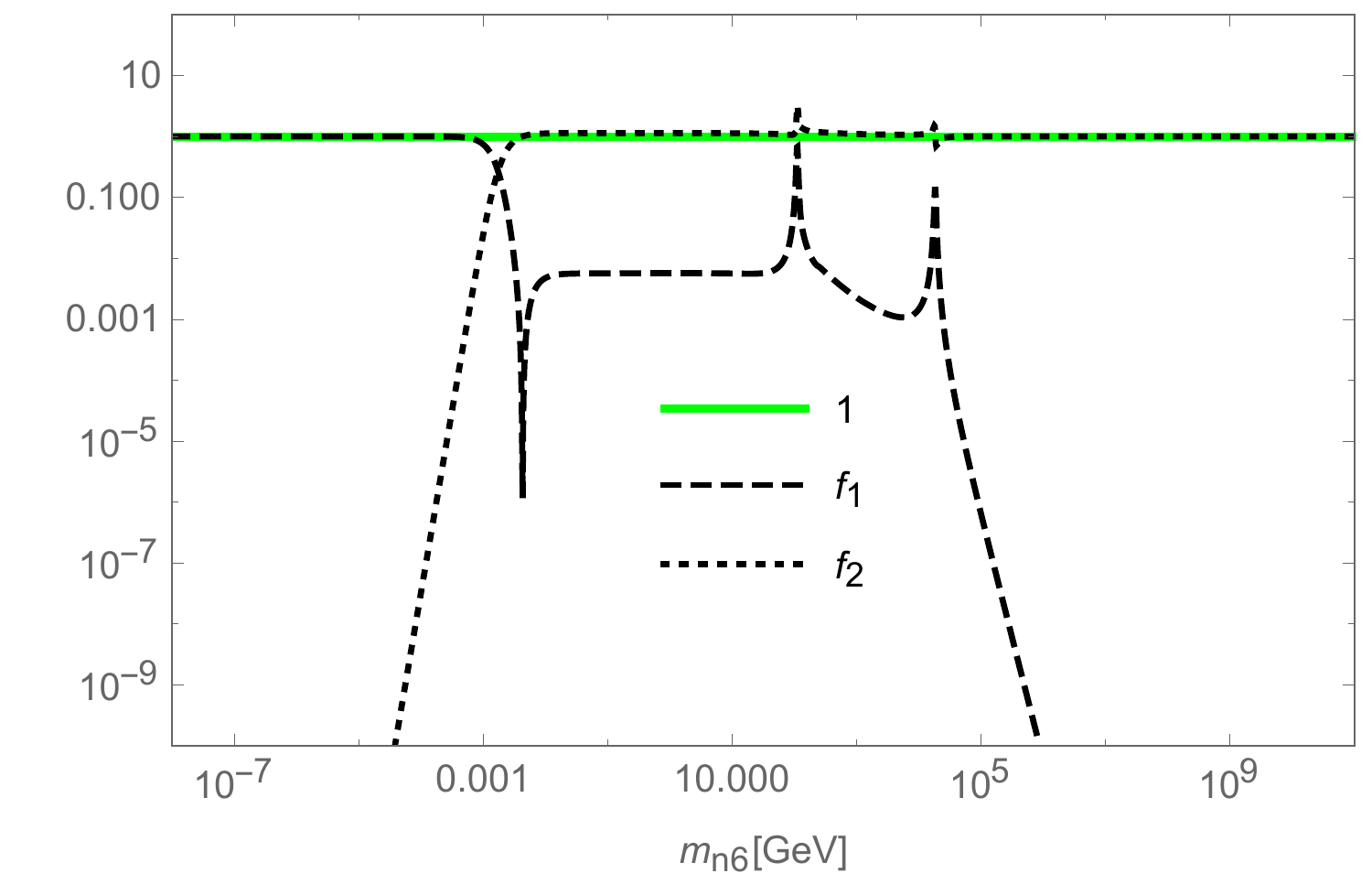}\\
	\end{tabular}
	\caption{Comparing different contributions to Br$(h\rightarrow \mu\tau)$ as functions of heaviest exotic neutrino mass $m_{n_6}$, where $3m_{n_4}=2m_{n_5}=m_{n_6}$, $f_1=(\mathrm{no\; terms\; with\; m^2_{n_i})/total}$, and $f_2=\mathrm{(only \;terms\; with\; m^2_{n_i})/total}$. }\label{NBrhmutau2}
\end{figure}
The largest component of the matrix $R$  satisfies $R\sim \mathcal{O}\left(\sqrt{\frac{|\hat{m}_{\nu}|}{m_{n_6}}}\right)$. As a result,  the  mixing matrix elements in $\Delta^{(a+c+d)}_{L,R}$ and $\Delta^{(b)}_{L,R}$ will results in the following factors: $U^{\nu*}_{a(I+3)}U^{\nu}_{b(I+3)}=|R_{aI}|^2\sim \frac{|\hat{m}_{\nu}|}{m_{n_6}}$. There are new factors in the $\Delta^{(b)}_{L,R}$:  $U^{\nu*}_{a(I+3)}U^{\nu}_{c(I+3)}U^{\nu*}_{c(J+3)}U^{\nu*}_{b(J+3)}\sim \frac{|\hat{m}_{\nu}|^2}{m^2_{n_6}}$. Hence the largest contribution to the total gives  $\Delta_{L,R}\sim m_{n_6}$ with very large $m_{n_6}$, implying  Br$(h\rightarrow\mu\tau)\sim m^2_{n_6}$.  The correlations between terms with and without factors $m^2_{n_i}$ are shown in the Fig. \ref{NBrhmutau2}.  Terms without factors $m_{n_i}^2$  are dominant with tiny $m_{n_6}$ but they are very suppressed with large $m_{n_6}$. 

The above discussions lead to new interesting results for  LFVHD predicted by the MSS model, which were not concerned previously: i) the Br can reach values of order $10^{-11}$ with large values of heavy neutrino masses satisfying the perurbative limit; ii) the Br enhances with increasing $m_{n_6}$ above $10^5$ GeV. In addition, the maximal Br$(h\rightarrow\mu\tau)$ reaches the values of $10^{-33}-10^{-32}$ with $m_{n_6}\in [10^2,10^4]$ GeV. We will show the relation between these  interesting values and  maximal values  of Br$(h\rightarrow\mu\tau)$ predicted by the ISS.

We realize that the property of Br$(h\rightarrow\mu\tau)\sim m^2_{n_6}$  agrees very well with the  approximate expression shown in  \cite{Apo}. In particular,  Br$(h\rightarrow\mu\tau)\sim m^4_{n_6}\times |F_N|^2$, where $F_N\sim R^2 \sim m^{-1}_{n_6}$ relating with active-heavy neutrino mixing elements in $U^{\nu}$.  We believe that large values of the Br predicted in \cite{Apo} arise  from the reason that recent neutrino oscillation data could not be applied at that times. The  numerical values of $F_N$ chosen in \cite{Apo} may keep large contributions  that should vanish because of  the GIM mechanism. 
 
Although the maximal Br of  LFVHD predicted by the MSS is much smaller than the prediction from the ISS model given in \cite{EArganda,iseesaw}, the behave of the curve presenting Br$(h\rightarrow\mu\tau)$ shown in Fig. \ref{NBrhmutau2} have the same form with Br$(h\rightarrow\mu\tau)$ calculated in the ISS. The reason is as follows. If  the exotic neutrino masses are fixed the same values  in the two models, $m_M=M_R=\mathrm{diag}(m_{n_4},m_{n_5},m_{n_6})$,  the important quantity making different contributions to LFVHD is the parametrization of $m_D$, see two Eqs. (\ref{fmD}) and  (\ref{mDiss}) for the MSS and ISS, respectively. 
\begin{figure}[ht]
	\centering 
	\begin{tabular}{cc}
		\includegraphics[width=8cm]{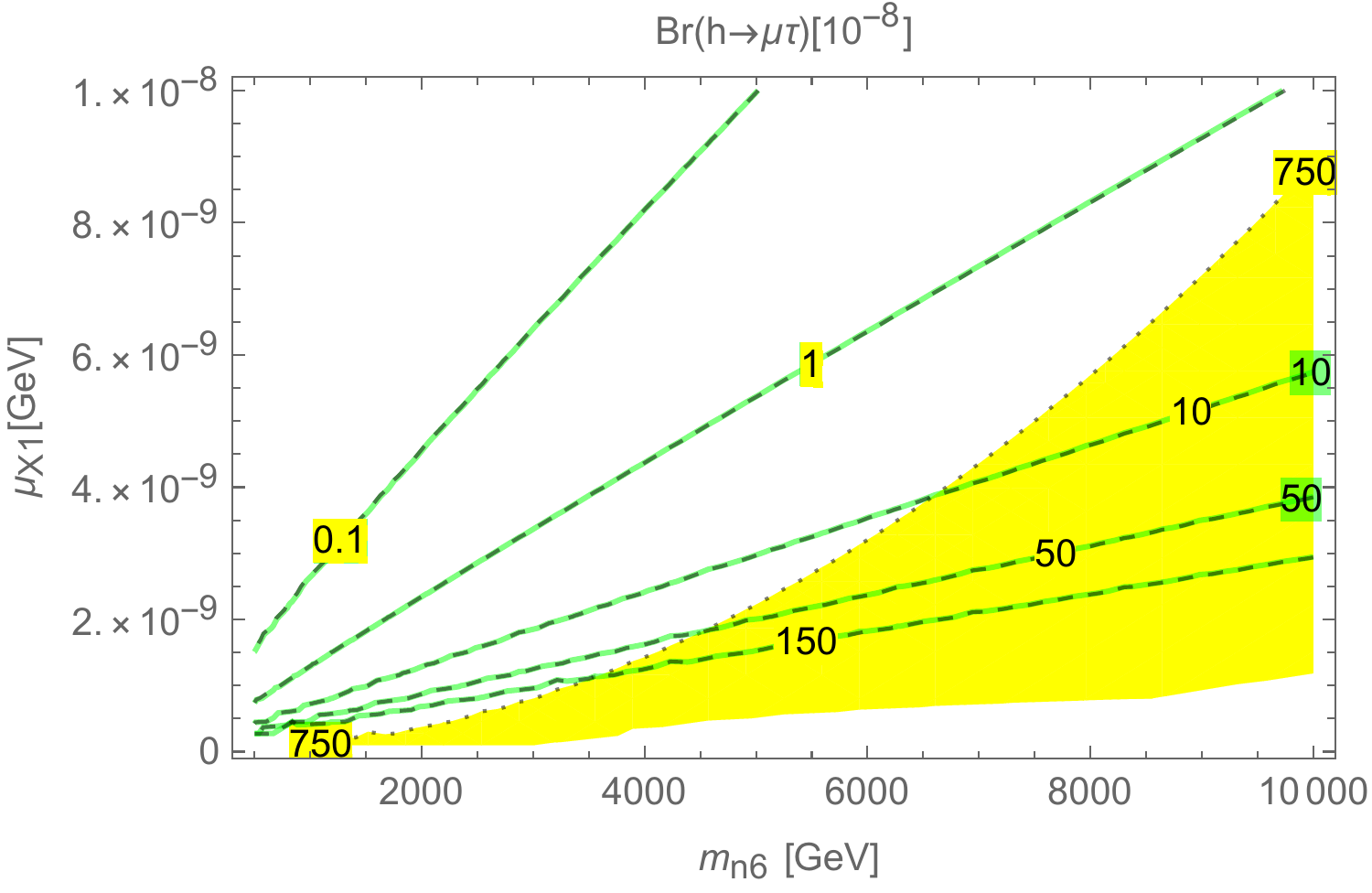}
& \includegraphics[width=8cm]{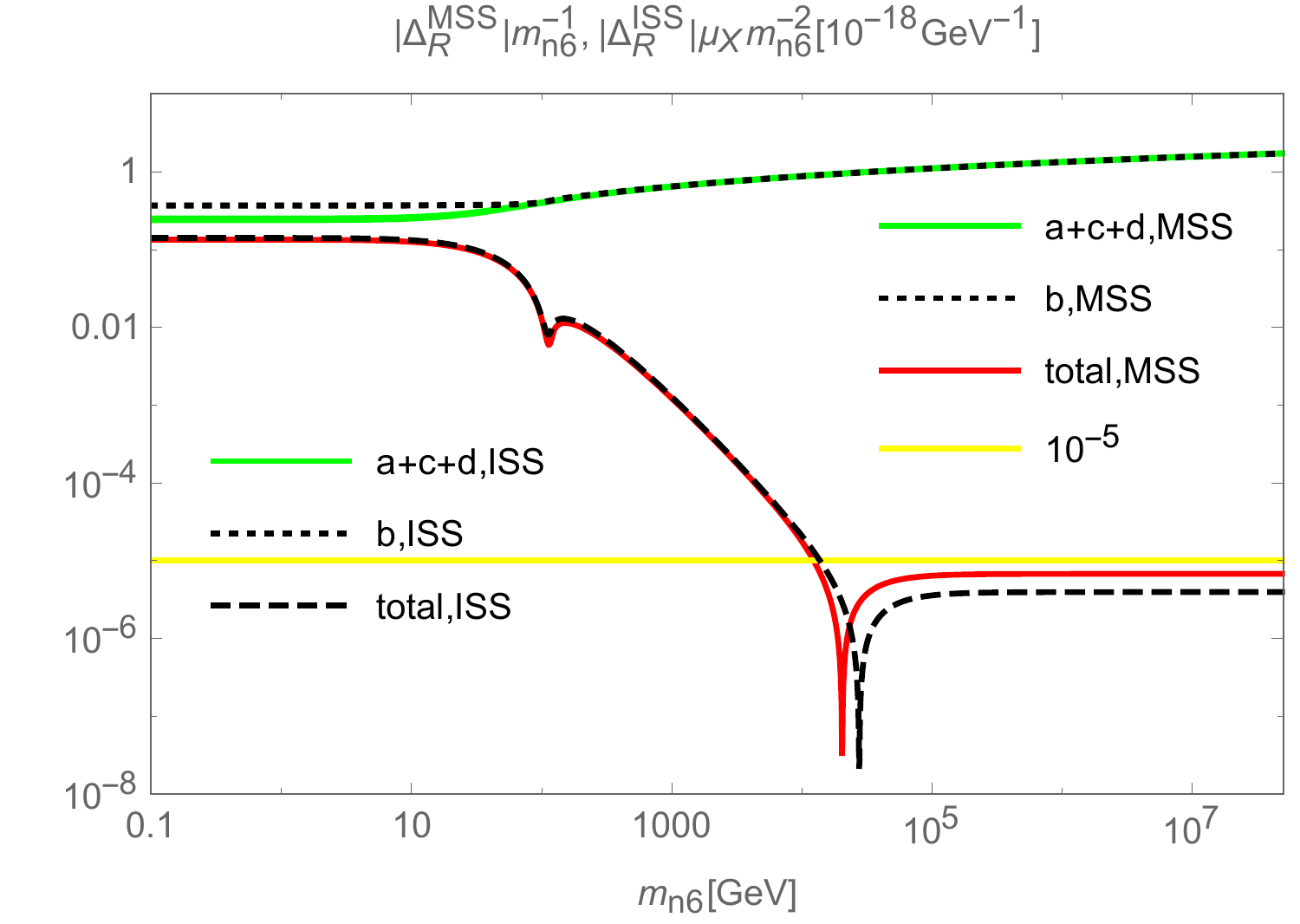}\\
	\end{tabular}
	\caption{Left panel: contour plot of Br$(h\rightarrow \mu\tau)$ and $|m_D|$ as functions of $m_{n_6}$ and $\mu_X$, predicted from ISS framework. The yellow region is excluded by large $|m_D|>174\sqrt{6\pi}$ GeV. Dashed black curves are from ISS prediction. Green curves obtained from modifying MSS. The right panel: a comparison between different contributions from  $|\Delta^{\mathrm{MSS}}_{R}|m^{-1}_{n_6}$ and  $|\Delta^{\mathrm{ISS}}_{R}|\mu_X m^{-2}_{n_6}$.}\label{NBrhmutau4}
\end{figure} 
This leads to the different structures of the $R$ matrices.   The largest components of $R$ in the MSS are   $R^{\mathrm{MSS}}_{a I}\sim \sqrt{\frac{|\hat{m}_{\nu}|}{|m_{n_6}|}}$ with $I>3$,  while those in the ISS are $R^{\mathrm{ISS}}_{aI}\sim \sqrt{\frac{|\hat{m}_{\nu}|}{\mu_X}}$.  Hence, in general the ISS mixing factors are larger than those of MSS a common factor $\sqrt{\frac{|\hat{m}_{n_6}|}{|\mu_X|}}$.  It makes the prediction of Br of LFVHD by the ISS  be  much larger than the prediction by the MSS, provided large $m_{n_6}$ but small $\mu_X$. Unlike the MSS, where mass scale $m_{n_6}$ can be as large as $\mathcal{O}(10^{15})$ GeV, values of $m_{n_6}$ in the ISS are constrained by relation (\ref{R1}), i.e. $m^2_{n_6} |\hat{m}_{\nu}|/\mu_X=|m_D|^2< 174^2\times 6\pi\, [\mathrm{GeV}]^2$. Hence, small $\mu_X$ will give small upper bounds of $m_{n_6}$, and large Br$(h\rightarrow\mu\tau)$ will depend complicatedly on these two parameters. The left panel of Fig. \ref{NBrhmutau4} shows possible values of Br$(h\rightarrow\mu\tau)$ in the allowed  regions of $\mu_{_X}$ and $m_{n_6}$. Our numerical results are well consistent with previous work \cite{iseesaw}. In addition, by adding a factor $\sqrt{\frac{|m_{n_6}|}{|\mu_X|}}$ into $R^{\mathrm{MSS}}$ and using the analytic expressions of $\Delta^{\mathrm{MSS}}_{L,R}$ we get a very consistent results of Br$(h\rightarrow\mu\tau)$ predicted by the ISS, see an illustration in the left panel of  Fig. \ref{NBrhmutau4}.  This confirms again the consistence of our calculation for LFVHD in the MSS and ISS. 
 
There is an interesting relation between two LFVHD amplitudes calculated in the two models, as drawn in the right panel of Fig. \ref{NBrhmutau4}. Here, $|\Delta^{\mathrm{ISS}}_{R}| \mu_Xm^{-2}_{n_6}$ and $|\Delta^{\mathrm{MSS}}_{R}|m^{-1}_{n_6}$ are considered as functions of $m_{n_6}$. We have checked numerically that  $|\Delta^{\mathrm{ISS}}_{R}| \mu_Xm^{-2}_{n_6}$ does not depend on $\mu_X$, and consistent with conclusion in  \cite{iseesaw}. It can be seen as follows.  The dependence of   $m_D$ and $R^{\mathrm{ISS}}$ on $M_R$ and $\mu_X$ can be separate into two parts. The first is the correlation between elements of these matrices in order to give correct experimental values of active neutrino data. And the second is the simple dependence on the scales of $m_{n_6}$ and $\mu_X$. In the ISS, $R^{\mathrm{ISS}}_{aI}=U^{\nu}_{a(I+3)}\sim \mu_X^{-1/2}$ and do not depend on $m_{n_6}$. Now, if we pay attention to the region with large $m_{n_6}$, the terms like $m_{n_i}^2 B_{0,1,2}$ are dominant contributions to $\Delta_{L,R}$ because of the factors $m_{n_i}^2$. As a result, $\Delta^{(a+c+d)}_{L,R}$ containing a factor $U^{\nu*}_{ai}U^{\nu}_{bi}\sim \mu^{-1}_X$ will give an overall factor $\mu_X^{-1}m^2_{n_6}$. Hence $\Delta^{(a+c+d)}_{L,R}\mu_X m^{-2}_{n_6}$ may be constant, following the property of $B$-functions. On the other hand,  $\Delta^{(b)}_{L,R}$ contains  $U^{\nu*}_{ai}U^{\nu*}_{cj} U^{\nu}_{ci}U^{\nu}_{bj}\sim \mu^{-1}_X$ or $\mu^{-2}_X$, depending on both indices $i$ and $j$ or only one larger than 3. Because both $\Delta^{(a+c+d)}_{L,R}$ and $\Delta^{(b)}_{L,R}$ are still divergent,  terms with $\mu^{-2}_{X}$ must vanish in order to guarantee a finite $\Delta_{L,R}$. This results in a common factor $\mu_X^{-1} m^2_{n_6}$ for $\Delta_{L,R}$.  In the right panel of Fig. \ref{NBrhmutau4}, values of $\mu_X m^{-2}_{n_6}\Delta^{(a+c+d)}_{L,R}$ and $\mu_X m^{-2}_{n_6}\Delta^{(b)}_{L,R}$ correspond to $\Delta_{\epsilon}=0$. But we checked numerically that $\mu_X^{-1} m^2_{n_6}\Delta_{L,R}$ is independent with $\Delta_{\epsilon}$. In addition, we can see that $\mu_X m^{-2}_{n_6}\Delta^{(a+c+d)}_{L,R}$ and $\mu_X m^{-2}_{n_6}\Delta^{(b)}_{L,R}$ always have opposite signs, which is consistent with the fact that divergences contained in them are really canceled. Two absolute  contributions from $\Delta^{(a+c+d)}_{L,R}$ and $\Delta^{(b)}_{L,R}$ are the same order, and  nearly degenerate with large $m_{n_6}$.  They start canceling strongly  each other from the electroweak range of $m_{n_6}$, giving a very small  $\mu_X m^{-2}_{n_6}\Delta_{L,R}$. It is $10^{-5}$ times smaller than values of $\mu_X m^{-2}_{n_6}\Delta^{(b)}_{L,R}$. 

The above discussion is the same for both models ISS and MSS, where $ m^{-1}_{n_6}\Delta_{L,R}$ is the function considered in the MSS. The numerical results are also shown in the right panel of the Fig. \ref{NBrhmutau4}. Consider a region $10\le m_{n_6}\le 10^{4}$ GeV, there is an equality that $m^{-1}_{n_6}\Delta^{\mathrm{MSS}}_{L,R}= \Delta^{\mathrm{ISS}}_{L,R}\mu_X m^{-2}_{n_6}$, implying $\mathrm{Br^{ISS}}(h\rightarrow\mu\tau)=\frac{m^2_{n_6}}{\mu^2_X}\mathrm{Br^{MSS}}(h\rightarrow\mu\tau)$. From previous discussion, where $\mathrm{Br^{MSS}}(h\rightarrow\mu\tau)\le 10^{-32}$, we can derive the maximal $\mathrm{Br^{ISS}}(h\rightarrow\mu\tau)\le 10^{-32}\times \mathcal{O}((10^4/10^{-9})^2)=\mathcal{O}(10^{-6})$. 

We can also estimate the maximal value of Br$(h\rightarrow\mu\tau)$ based on the numerical result shown in Fig. \ref{NBrhmutau4}. If $m_{n_6}\geq 10^{5}$ GeV, we have $ \Delta_{R}\simeq  10^{-24}\mu_X^{-1} m^{2}_{n_6}$, where small $\Delta_{L}$ is ignored. Equivalently, we have  Br$(h\rightarrow\mu\tau)\simeq  10^{-45}\mu_X^{-2} m^{4}_{n_6}$. The condition of perturbative limit gives $m^2_{n_6}\times 5\times 10^{-11}/\mu_{X}=|m_D|^2\leq 174^2\times 6\pi$, leading to $ \mu_X^{-2} m^{4}_{n_6}\le \mathcal{O}(10^{36})$. Hence in the region of lagre $m_{n_6}\ge 10^5$ GeV, Br$(h\rightarrow\mu\tau)$ can reach maximal value of $\mathcal{O}(10^{-9})$. If  $m_{n_6}< 10^{5}$ GeV, the allowed region in the left panel of Fig. \ref{NBrhmutau4} shows that Br$(h\rightarrow\mu\tau)$ can reach values of $\mathcal{O}(10^{-7})$ only if $m_{n_6}$ is few TeV, $\mu_X$ is order of $10^{-9}$ GeV, and $m_D$ gets values very close to the perturbative limit.

\section{\label{Con}Conclusion} 
In this work, the LFVHD in the MSS and ISS models have been discussed where we have focused on  new aspects that were not shown in previous works. We calculated the amplitude of the LFVHD using new analytical expressions of PV-functions discussed recently. From this we have checked  the consistence of our results in many different ways: comparing them with results of  previous works, calculating in two gauges of unitary and 't Hooft-Feynman, checking analytically the divergent cancellation of the total amplitude.  In the MSS framework, we investigated numerically the Br$(h\rightarrow \mu\tau)$ in the valid and large range of exotic neutrino mass scale, from $10^{-10}$ GeV to $10^{16}$ GeV. When applying the Casas-Ibarra parameterization to Yukwa couplings of heavy neutrinos, we found a new result that Br$(h\rightarrow\mu\tau)\sim m^2_{n_6}$ with large  $m_{n_6}$, because the mixing matrix elements affecting mostly the LFVHD amplitude by factors of $m_{n_6}^{-1/2}$. But in the valid region of perturbative requiring $m_{n_6}<10^{16}$ GeV, the Br$(h\rightarrow\mu\tau)$ reaches maximal values of $\mathcal{O}(10^{-11})$, still far from the recent experimental consideration. Anyway, this may be a hint to improve the MSS to more relevant models predicting higher values of Br$(h\rightarrow\mu\tau)$, for example the ISS. In this model, the largest mixing factors contributing  to LFVHD amplitude do not depend on the exotic neutrino mass scale $m_{n_6}$ but consist of a factor $\mu_X^{-1}$. Hence, if two models have the same neutrino mass scale, and  the neutrino mixing matrices  obey the Casas-Ibarra parameterization, there will be a very simple relation that  $\mathrm{BR^{ISS}}(h\rightarrow\mu\tau)/\mathrm{BR^{MSS}}(h\rightarrow\mu\tau) \simeq m^2_{n_6}\mu^{-2}_X$. This explains why the signal of LFVHD in the ISS is extremely significant than that in MSS. But the perturbative condition does not allow both large $m_{n_6}$ and small $\mu_X$, which can predict large Br$(h\rightarrow\mu\tau)$. Hence, maximal Br$(h\rightarrow\mu\tau)$ is still $\mathcal{O}(10^{-7})$ with few TeV of heavy neutrino mass scale. Our discussion on LFVHD of the MSS suggests that  Br$(h\rightarrow\mu\tau)$ may be large in the extended versions of the MSS which allow very large $m_{n_6}$. Finally, although we presented here a different way to calculate the LFVHD, our numerical results for the ISS are well consistent with those noted in previous works \cite{iseesaw,e1612.09290}.  

\section*{Acknowledgments}
  LTH thanks Professor Thomas Hahn for useful discussions  on Looptools. We are especially  thankful Dr. Ernesto  Arganda and authors of Refs. \cite{EArganda,iseesaw,e1612.09290} for important comments to correct our calculations and help us understand more deeply the LFVHD in the ISS. We thank Dr. Julien Baglio and Dr. LE Duc Ninh for helpful discussions.  This research is funded by Ministry of  Education  and Training under grant number B2017\_SP2\_06.
\appendix
\section{\label{PVfunction}One loop Passarino-Veltman functions}
Calculation in this section relates with one-loop diagrams
 in the Fig. \ref{Feydia1}.  The analytic expressions of the PV-functions are given in \cite{LFVHDUgauge}and  they were derived from the general forms given in \cite{Hooft}, using only the conditions of very small masses of tau and muon. They are consistent with \cite{bardin}. The denominators of the propagators  are denoted as  $D_0=k^2-M_0^2+i\delta$, $D_1=(k-p_1)^2-M_{1}^2+i\delta$ and $D_2=(k+p_2)^2-M_2^2+i\delta$, where $\delta$ is  infinitesimally a positive real quantity. The scalar integrals are defined as
 \bea
 B^{(i)}_0 &\equiv&\frac{\left(2\pi\mu\right)^{4-D}}{i\pi^2}\int \frac{d^D k}{D_0D_i},\quad 
   B^{(12)}_0 \equiv \frac{\left(2\pi\mu\right)^{4-D}}{i\pi^2}\int \frac{d^D k}{D_1D_2},\crn
 C_0 &\equiv&   C_{0}(M_0,M_1,M_2) =\frac{1}{i\pi^2}\int \frac{d^4 k}{D_0D_1D_2},\nn
 \label{scalrInte}\eea
 where $i=1,2$.
   In addition, $D=4-2\epsilon \leq 4$ is the dimension of the integral;  $M_0,~M_1,~M_2$ are masses of virtual particles in the loop. The momenta satisfy conditions: $p^2_1=m^2_{1},~p^2_2=m^2_{2}$ and $(p_1+p_2)^2=m^2_{h}$. In this work, $m_h$ is the SM-like Higgs mass, $m_{1,2}$ are lepton masses. The tensor integrals are
 \bea
 B^{\mu}(p_i;M_0,M_i)&=& \frac{\left(2\pi\mu\right)^{4-D}}{i\pi^2}\int \frac{d^D k\times
k^{\mu}}{D_0D_i}\equiv B^{(i)}_1p^{\mu}_i,\crn
C^{\mu} &=&C^{\mu}(M_0,M_1,M_2)=\frac{1}{i\pi^2}\int \frac{d^4 k\times k^{\mu}}{D_0D_1D_2}\equiv  C_1 p_1^{\mu}+C_2 p_2^{\mu}.\nn  \label{oneloopin1}\eea
  The PV functions are $B^{(i)}_{0,1}$, $B^{(12)}_{0}$ and $C_{0,1,2}$. The  functions $C_{0,1,2}$ are finite while the remains are divergent. We define the common divergent part as $\Delta_{\epsilon}\equiv \frac{1}{\epsilon}+\ln4\pi-\gamma_E+\ln\mu^2$ where $\gamma_E$ is the  Euler constant.  Then the divergent parts of the above scalar factors are $\mathrm{Div}[B^{(i)}_0] = \mathrm{Div}[B^{(12)}_0]= \Delta_{\epsilon}$, and $\mathrm{Div}[B^{(1)}_1]=- \mathrm{Div}[B^{(2)}_1]= \frac{1}{2}\Delta_{\epsilon}$.

  For  simplicity  in calculation we  use  approximative forms of PV functions where $p_1^2,p_2^2\rightarrow 0$.  The  function $C_0$ was given in \cite{LFVHDUgauge} consistent with that discussed on \cite{bardin}, namely
 \be  C_0=\frac{1}{m_h^2}\left[R_0(x_0,x_1)+ R_0(x_0,x_2)-R_0(x_0,x_3)\right] ,\nn \label{C0fomula1}\ee
 where
 $ R_0(x_0,x_i) \equiv Li_2(\frac{x_0}{x_0-x_i})- Li_2(\frac{x_0-1}{x_0-x_i})$,  $Li_2(z)$ is the di-logarithm function;  $x_{1,2}$ are solutions of the equation $x^2-\left(\frac{m_h^2-M_1^2+M_2^2}{m_h^2}\right)x+\frac{M_2^2-i\delta}{m_h^2}=0$;  $ x_0=\frac{M_2^2-M_0^2}{m_h^2}$; and $x_3=\frac{-M_0^2+i\delta}{M_1^2-M_0^2}$.

Based on \cite{denner}, the $B$-functions with small absolute values of  external momenta can be written in stable forms in numerical computations.  Defining $y_{ij}$ ($i,j=1,2$) are solutions of the equation $ y^2p^2-y(p^2_i+M^2_i-M^2_0)+M_i^2-i\delta=0$.  New functions $f_{n}(y)$ are defined as follows,
\be  f_n(y)\equiv (n+1) \int_0^1 dt~ t^n\ln\left(1-\frac{t}{y}\right), \nn\label{fnx} \ee
so that they can be evaluated numerically stable way by choosing
\bea  f_n(x)=\left\{
\begin{array}{cc}
	\left (1-y^{n+1}\right)\ln\frac{y-1}{y}-\sum_{l=0}^{n} \frac{y^{n-l}}{l+1} & \mathrm{ if} ~|y|<10, \\
	\ln\left(1-\frac{1}{y}\right)+\sum_{l=n+1}^{\infty}\frac{y^{n-l}}{l+1}&\mathrm{ if }~|y|\geq10 .\\
\end{array}
\right. \nn\label{nbfun}
\eea
The $B$-functions now can be expressed in terms of $ f_n(y)$, namely  
\bea B_0^{(i)}&=& \Delta_{\epsilon}-\ln M_i^2- \sum_{j=1}^2f_0(y_{ij}),\crn
 B_1^{(i)} &=&(-1)^i \left[\frac{1}{2} \left( \Delta_{\epsilon}-\ln M_i^2\right)-  \sum_{k=1}^2f_0(y_{ij})+ \frac{1}{2}\sum_{k=1}^2f_1(y_{ij})\right].\nn\label{B0f}\eea
Finally, the $B^{(12)}_0$ and $C_{1,2}$ functions are determined as follows, 
 \bea 
B^{(12)}_0&=& \Delta_{\epsilon} -\ln M_1^2 +2 + \sum_{k=1}^2 x_k\ln\left(1-\frac{1}{x_k}\right),\crn
 C_1 &=& \frac{1}{m_h^2}   \left[B^{(1)}_0 -B_0^{(12)}+(M_2^2-M_0^2)C_0\right],\;
  C_2 =  -\frac{1}{m_h^2}   \left[B^{(2)}_0 -B_0^{(12)}+(M_1^2-M_0^2)C_0\right]. \nn\eea
  
In our work above use the following notations, $m_1\equiv m_a$, $m_2\equiv m_b$, $p_1\equiv p_a$ and $p_2\equiv p_b$.  
\section{\label{match}Matching with notations in previous works}
This section will show the equivalence given in (\ref{twonotation}). We recall  notations used in \cite{EArganda,iseesaw,e1612.09290} as follows. The external momenta are $p'_1$,$(-p_2')$, and $p_3'$  for  ingoing Higgs boson, outgoing leptons $e_a$ and $e_b$,respectively. The prime is used to distinguish from the notions that were used in our work, especially those given in Sec. \ref{PVfunction}. Three denominators of the propagators are $D'_0= k^2-m_1^2$, $D'_1= (k+p'_2)^2-m_2^2$ and $D'_2= (k+p'_1+p'_2)^2-m_3^2$. The one-lopp-three-point functions are defined as,
\bea  \int \frac{d^4k}{(2\pi)^4}\times \frac{\left\{1,k^{\mu} \right\}}{D'_0D'_1D'_2} &=& \frac{i}{16\pi^2} \left\{C'_0,\, C'_{\mu}= C_{11}p'_{2\mu}+ C_{12}p'_{1\mu}\right\}.\label{NPVf}\eea

The equivalence between above notations with those given in Sec. \ref{PVfunction} are $p'_1=p_1+p_2$, $ p'_2=-p_1$, $m_{1,2,3}=M_{0,1,2}$. As a result, we get $D'_{0,1,2}=D_{0,1,2}$, leading to $C'_0=C_0$ and $C'_{\mu}=C_{\mu}$.  But the scalar factors  $C_{11,12}$ and $C_{1,2}$ are different, namely $C'_{\mu}= C_{11}(-p_{1\mu})+ C_{12}(p_{1\mu}+ p_{2\mu})=(C_{12}-C_{11})p_{1\mu} + C_{12}p_{2\mu}$. Matching this with definition of $C_{\mu}$ defined in Sec. \ref{PVfunction}. We obtain  the equivalence for $C_{1,2}$ in (\ref{twonotation}).  Other $B$-functions is proved easily so we omit here.  

\section{\label{Diagramb} Form factors  in unitary gauge for LFVHD} 
The contribution from diagram in Fig. \ref{Feydia1}b) to the LFVHD amplitude is
{\small
\bea i \mathcal{M}_{(b)}&=& \int \frac{d^4k}{(2\pi)^4}\times \bar{u}_a \left( \frac{ig}{\sqrt{2}}U^{\nu*}_{ai}\gamma^{\mu}P_L\right) \frac{i\left[(-k\!\!\!\slash+p_a \!\!\!\!\!\slash\,)+m_{n_i}\right]}{D_1}\crn
&\times& \left[ \frac{-ig}{2m_W}\sum_{c=1}^{3}C_{ij}\left(m_{n_i}P_L + m_{n_j}P_R \right)+ C^*_{ij}\left(m_{n_j}P_L + m_{n_i}P_R \right)\right] \crn
&\times& \frac{i\left[-(k\!\!\!\slash+p_b\!\!\!\!\!\slash \,)+m_{n_j}\right]}{D_2}\times \left( \frac{ig}{\sqrt{2}}U^{\nu}_{bj}\gamma^{\nu}P_L\right) v_b\times \frac{-i}{D_0} \times \left( g_{\mu\nu}-\frac{k_{\mu}k_{\nu}}{m^2_W}\right) \crn
&=& \frac{-g^3}{4m_W} \sum_{i,j=1}^{K+3}\sum_{c=1}^{3}U^{\nu*}_{ai}U^{\nu}_{bj}\times \int \frac{d^4k}{(2\pi)^4} \frac{1}{D_0D_1D_2}\times \left(g_{\mu\nu}-\frac{k_{\mu}k_{\nu}}{m^2_W}\right)\crn
&\times& \bar{u}_a\gamma^{\mu}P_L\left[(-k\!\!\!\slash+p\!\!\slash_a)+m_{n_i}\right] \left[ C_{ij}\left(m_{n_i}P_L + m_{n_j}P_R \right)  \right.\crn
&+& 
\left. C^*_{ij} \left(m_{n_j}P_L + m_{n_i}P_R \right) \right]   \left[-(k\!\!\!\slash+p\!\!\slash_b)+m_{n_j}\right] \gamma^{\nu}P_Lv_b. \nn \label{Mefvv1}\eea
}
The final result is 
 \bea i \mathcal{M}_{(b)} &=& \frac{i}{16\pi^2}\times  \frac{-g^3}{4m^3_W} \sum_{i,j=1}^{K+3}\sum_{c=1}^{3}U^{\nu*}_{ai}U^{\nu}_{bj} \crn &\times&  \left\{ m_a [\overline{u_a}P_Lv_b]\left[ C_{ij} \left( m^2_{n_i}B^{(1)}_1 +m^2_{n_j}B^{(12)}_0 +2 \left[m^2_{n_i} m^2_{n_j} +m_W^2(m^2_{n_i}+m^2_{n_j}) \right]C_1\right. \right.\right.\crn &&\left.\left.\left.-(m^2_{n_i}m^2_b+m^2_{n_j}m^2_a)C_1 -m^2_{n_j}m^2_W C_0 \right)\right.\right.\crn
 &+&\left.\left. C^*_{ij} m_im_j\left( B^{(12)}_0+ B^{(1)}_1-m_W^2 C_0 +\left[4 m_W^2+m_{n_i}^2+ m^2_{n_j}-m_a^2-m_b^2\right]C_1\right)  \right] \right. \crn
 &+&\left. m_b [\overline{u_a}P_Rv_b]\left[C_{ij} \left( -m^2_{n_j}B^{(2)}_1  +m^2_{n_i}B^{(12)}_0  -2 \left[m^2_{n_i} m^2_{n_j} +m_W^2(m^2_{n_i}+m^2_{n_j}) \right]C_2\right.\right.\right.\crn&&\left.\left.\left.+(m^2_{n_i}m^2_b+m^2_{n_j}m^2_a)C_2 -m^2_{n_i}m^2_W C_0\right) \right.\right.\crn
 &+&\left.\left.C^*_{ij} m_im_j\left( B^{(12)}_0- B^{(2)}_1-m_W^2 C_0 -\left[4 m_W^2+m_{n_i}^2+ m^2_{n_j}-m_a^2-m_b^2\right]C_2\right)\right]\right\}.\nn  \label{Mefvv2}\eea

\end{document}